\documentclass[10pt,prd,aps,superscriptaddress,floatfix,nofootinbib,twocolumn,eqsecnum]{revtex4-2}
\pdfoutput=1
\usepackage[utf8]{inputenc}
\usepackage{amsfonts}
\usepackage{amsmath}
\usepackage{amssymb}
\usepackage{bm}
\usepackage{dcolumn}
\usepackage{subfigure}
\usepackage{graphicx}
\usepackage[english]{babel}
\usepackage{multirow}

\usepackage{empheq}
\usepackage{latexsym}
\usepackage{placeins}
\usepackage[normalem]{ulem}
\usepackage[T1]{fontenc}
\usepackage{booktabs}
\usepackage[table]{xcolor}
\usepackage[unicode]{hyperref}
\usepackage{changes}
\usepackage{makecell}
\usepackage{ragged2e}

\usepackage{caption}
\begin{document}

\title{Polymerized spacetime dynamics with multifield source: Unraveling the preinflationary Universe}

\author{Divya Gupta}\email{divya920818@gmail.com}
\affiliation{Centre for Theoretical Physics and Natural Philosophy, Nakhonsawan Studiorum for Advanced Studies,
Mahidol University, Nakhonsawan 60130, Thailand}

\author{Manabendra Sharma} \thanks{Corresponding author}
\email{sharma.man@mahidol.ac.th}
\affiliation{Centre for Theoretical Physics and Natural Philosophy, Nakhonsawan Studiorum for Advanced Studies,
Mahidol University, Nakhonsawan 60130, Thailand}

\author{Gustavo S. Vicente}
\email{gustavo@fat.uerj.br}
\affiliation{Faculdade de Tecnologia, Universidade do Estado do Rio de Janeiro, 27537-000 Resende, RJ, Brazil}

\author{Rudnei O. Ramos}
\email{rudnei@uerj.br}
\affiliation{Departamento de F\'{\i}sica Te\'orica, Universidade do Estado do Rio de Janeiro, 20550-013 Rio de Janeiro, RJ, Brazil}

\author{Anzhong Wang}
\email{Anzhong$\_$Wang@baylor.edu}
\affiliation{GCAP-CASPER,  Department of Physics $\&$ Astronomy, Baylor University, Waco, Texas 76798-7316, USA}

\begin{abstract}\label{sec:abstract}

We study a multifield model in loop quantum cosmology for a maximally symmetric spacetime governed by the Einstein-Hilbert action minimally coupled to scalar fields. Using a Legendre transformation, we formulate the Hamiltonian dynamics in canonically equivalent geometrodynamical and Yang-Mills-type representations, incorporating nontrivial couplings through a geometric structure on the multifield configuration space. Implementing the $\bar{\mu}$-scheme polymerization, we obtain the loop-quantum-corrected Friedmann equations. By focusing on the two-field models as an example, we analyze the effective dynamics for specific potentials. The \textit{quantum bouncing, transition, and slow roll inflationary} phases are investigated numerically, and viability of the models is assessed by evaluating the number of e-folds during the inflationary phase for certain given initial conditions. The global behavior of the background evolution is further examined through linear stability and dynamical-systems analyses.

\end{abstract}

\maketitle 


\section{Introduction}\label{sec:intro}

Loop quantum cosmology (LQC), obtained by applying LQG techniques to symmetry-reduced spacetimes, provides an effective framework for investigating high-curvature regimes of the early Universe~\cite{Ashtekar:2011ni, Agullo:2023rqq,Li:2023dwy}. 
A key prediction of LQC is the resolution of the classical big bang singularity, which is replaced by a nonsingular quantum bounce~\cite{Ashtekar:2003hd, Bojowald:2005epg, Ashtekar:2008zu, Banerjee:2011qu, Ashtekar:2016ecx, Bojowald:1999tr,Ashtekar:2006wn, Ashtekar:2007em, Bojowald:2001xe, Singh:2009mz,Singh:2014fsy}. This phenomenon arises purely from quantum geometric effects, without invoking exotic matter. 
In this scenario, classical divergences in energy density and curvature are naturally regulated by the existence of a minimal, nonzero eigenvalue for the area operator—a consequence of the underlying discrete geometry\footnote{\textcolor{black}{Note that while the minimal area nonzero eigenvalue in LQG leads to a maximum energy density $\rho_c$ and resolves strong curvature singularities [e.g., the big bang singularity in isotropic FRW models], it is known that weak singularities---such as sudden or type-IV singularities---are not automatically avoided in LQC~\cite{Singh:2009mz,Saini:2018tto}. Our analysis presented in this paper mainly focuses on the preinflationary dynamics where strong singularities are resolved.}}. 
The robustness of singularity resolution in LQC extends beyond Einstein-Hilbert gravity (general relativity [GR]) and has been established for nontrivial spacetimes, different cosmologies including the ones exhibiting anisotropic cosmologies~\cite{Ashtekar:2009vc, Sharma:2019okc, Sharma:2023abm, Chiou:2007mg, Martin-Benito:2008dfr, Ashtekar:2009um, Wilson-Ewing:2010lkm, Gupt:2012vi, Singh:2013ava, Singh:2016jsa, Brizuela:2009nk, Agullo:2020iqv,Agullo:2022klq,Garcia-Quismondo:2019dwa,Garcia-Quismondo:2019kav,Gan:2025uvt}.
 
Interestingly, such an effective setup naturally leads to a phase of slow roll inflation following the quantum bounce~\cite{Ashtekar:2011rm, Li:2019ipm, Linsefors:2013cd, Barboza:2020jux, Barboza:2022hng, Corichi:2010zp}. 
Over the past two decades, a substantial body of work has explored this single-field framework in LQC, covering a variety of inflationary potentials and analyzing their implications for early Universe dynamics---see, for example, Refs.~\cite{Sharma:2018vnv, Sharma:2019okc, Mielczarek:2009zw,Graef:2018ulg,Benetti:2019kgw,Levy:2024naz} and references therein---but only a limited number of works have considered multifield systems~\cite{Ranken:2012hp}. However, to the best of our knowledge, multifield scenarios, particularly with nontrivial field-space geometry have not been extensively investigated within the effective dynamics of LQC. In this work, we aim to address this gap by studying the coupled dynamics of multiple scalar fields in a holonomy-corrected cosmological background.

Scalar field theory has long served as a foundational tool for constructing cosmological models, especially those aimed at realizing inflationary dynamics. Moreover, the dynamical richness increases when the nontrivial metric of multifield space and cross-coupling terms between scalar fields are introduced~\cite{Paliathanasis:2022luh, Chen:2022ccf, DeAngelis:2023fdu, Mizuno:2017idt, Brown:2017osf}.

Traditional slow roll models can face fine-tuning issues~\cite{Lyth:1998xn}, while purely kinematic models such as hyperinflation may not terminate naturally~\cite{Brown:2017osf}. 
This is where Linde's hybrid inflation paradigm becomes relevant~\cite{Linde:1993cn}. 
By coupling the inflaton to a ``waterfall'' field, inflation ends abruptly via a symmetry-breaking instability when the inflaton crosses a critical value, a mechanism controlled by potential dynamics rather than the gradual failure of slow roll~\cite{Linde:1993cn}. 
Subsequent studies have shown that the successful trajectories in hybrid models form a fractal attractor in field space, making inflation generic for a broad range of initial conditions~\cite{Ramos:2001zw}. Moreover, hybrid inflation has been successfully embedded within supersymmetric and supergravity frameworks, improving their alignment with particle physics and observational constraints~\cite{Lyth:1998xn, Linde:1997sj}.

In LQC, where quantum gravitational corrections such as inverse volume and holonomy effects significantly alter the early Universe's dynamics~\cite{Cailleteau:2013kqa, Grain:2009cj, Chiou:2010nd, Wilson-Ewing:2011gnq, Corichi:2013usa, Bojowald:2005epg}, it becomes essential to understand how these multifield inflationary mechanisms behave. 
Since LQC modifies Friedmann dynamics, the collective behavior of scalar fields could offer new insights into early Universe acceleration, attractor behavior, and potentially distinguishable phenomenological signatures from standard inflationary cosmology.

In this paper, our main interest is to understand the background dynamics of multifield scalar models in the context of LQC. We will focus on two representative multifield inflationary scenarios in this paper: the hybrid inflation potential and a string-inspired potential model. We analyze the background dynamics of both models within the framework of LQC and assess their viability, i.e., for producing sufficient inflation, when initial conditions are set at the bounce instant. 

This paper is organized as follows: in Sec.~\ref{sec:multifielddyn} we briefly review both the classical and quantum formulations appropriate for the multifield models, leading to their background dynamics in LQC. 
In Sec.~\ref{models}, we present the
two representative multifield models we study in this work: a two-field model as used hybrid inflation and a model inspired by string theory, which displays kinetic interaction between the fields. The numerical analysis of the hybrid model is presented in Sec.~\ref{sec:bg_hm}, while in Sec.~\ref{sec:bg_sm}, the detailed analysis for the second model is presented.
{}Finally, our conclusions and future perspectives are summarized in Sec.~\ref{sec:conclusion}. One Appendix is included, where we present some of the technical details.

\section{Dynamics for Multifield Scenario}
\label{sec:multifielddyn}

In this section, we first review  the classical formulation of a generic minimally coupled multifield scenario in connection dynamics for Friedmann–Lemaître–Robertson–Walker (FLRW) spacetime. Then, at the end, we present its quantization leading to the LQC dynamics. 

\subsection{Classical formulation}
\label{subsec:classicaldynamics}

We begin with the Einstein-Hilbert action
\begin{equation}
\label{eq:EH_Action_multifield}
S = \int d^4x \sqrt{-g} \left[ \frac{1}{2 \kappa} R - \frac{1}{2} G_{IJ}(\phi^K) \partial_\mu \phi^I \partial^\mu \phi^J - V(\phi^K)\right],
\end{equation}
where $\kappa=8\pi/m_{\rm Pl}^2$, with $m_{\rm Pl}$ is the Planck mass, $g$ is the determinant of metric $g_{\mu\nu}$, $R$ is the scalar curvature term obtained from the trace of the Ricci tensor $R_{\mu \nu}$, $G_{IJ}(\phi^K)$ is the metric living in the space of  scalar fields~\cite{Iacconi:2023slv}, and $V(\phi^K)$ is the multifield potential.
This represents a special case of noncanonical scalar field scenario~\cite{Nandi:2016pfr}, where a nontrivial choice of $G_{IJ}$ introduces a non-Euclidean field-space geometry and hence non-canonical kinetic term, with $G_{IJ}=\delta_{IJ}$ representing the canonical limit.

In conformity with the cosmological principle, we invoke the FLRW metric: 
\begin{eqnarray}
ds^2= -N^2(t)\; dt^2 + a(t)^2 \left( dx^2 + dy^2 + dz^2 \right),\label{eq:FLRWmetric}
\end{eqnarray}
to describe the homogeneous and isotropic Universe, where $N(t)$ is the lapse function. Throughout this work, we consider the gauge $N(t)=1$.
 The background dynamics are governed by the Einstein field equations and Klein-Gordon equations for each field $\phi^K$~\cite{Iacconi:2023slv}, as follows: 
\begin{subequations}\label{eq:multifield_eqns}
\begin{align}
3 H^2 = \kappa \left(\frac{\dot{\sigma}^2}{2} + V(\phi^K)\right)\label{eq:multifield_FD},\\
\dot{H} = -\frac{\kappa \dot{\sigma}^2}{2}\label{eq:multifield_RC},\\
D_t \dot{\phi}^K + 3 H \dot{\phi}^K + G^{IK} V_{,I}(\phi^I)\ = 0\label{eq:multifield_KG},
\end{align}
\end{subequations}
where the overdot means derivative with respect to cosmic time,
$H \equiv \frac{\dot a}{a}$ is the Hubble parameter; $\dot{\sigma}^2/2$ is the total kinetic energy, where $\sigma$ is the “adiabatic field” which represents the path length along the trajectory, such that $\dot{\sigma^2} = G_{IJ} \dot{\phi}^I \dot{\phi}^J$, $V_{,I}(\phi^I) \equiv \frac{\partial V}{\partial \phi^I}$; and $D_t$ is the covariant derivative in the multifield space, defined as $D_t A^I \equiv \dot{A^I} + \Gamma^I_{JK} \dot{\phi^J} A^K$, where $\Gamma^I_{JK}$ are the Christoffel symbols compatible with the multifield metric $G_{IJ}$~\cite{Iacconi:2023slv}.\footnote{The field-space metric $G_{IJ}$ is symmetric, 
$G_{IJ}=G_{JI}$~\cite{Iacconi:2023slv}, and admits an inverse $G^{IJ}$ satisfying 
$G^{IK}G_{KJ}=\delta^{I}_{\;J}$.}
From the resultant kinetic term $\dot{\sigma^2}/2$, 
the total energy density and pressure, respectively, read as
\begin{align}
\rho&=\frac{\dot{\sigma}^2}{2} + V(\phi^K),\label{eq:multifield_rho}\\
P&=\frac{\dot{\sigma}^2}{2} - V(\phi^K).\label{eq:multifield_P}
\end{align}
From the Ricci scalar $R=6 \left( \frac{\ddot{a}}{a} + \frac{\dot{a}^2}{a^2} \right)$ and using $\sqrt{-g}=a^3$, the Lagrangian density $\mathcal{L}$ is reduced to
\begin{eqnarray}
\mathcal{L}= -\frac{3 \dot{a}^2 a}{\kappa}  + \frac{a^3}{2} G_{IJ} \dot{\phi^I} \dot{\phi^J}  - a^3 V(\phi^K).
\label{eq:multifield_Lagrangian}
\end{eqnarray}
The configuration space of the Lagrangian density consists of $(a, \phi^K)$, with K corresponding to the number of fields in the multifield space. In order to obtain the quantum-corrected dynamics via LQG technique, the present theory must be canonically formulated. 
For this, we need to transform the configuration space to phase space, i.e., $(a, \phi^K)$ $\rightarrow (a, p_a, \phi^K, p_{\phi^K})$, representing the symmetry-reduced geometrodynamic phase space. The generalized momentum corresponding to the geometric degree of freedom is
\begin{eqnarray}
p_a \equiv\frac{\delta \mathcal{L}}{\delta \dot{a}} = -\frac{6}{\kappa} a\dot{a},\label{eq:momenta_a}
\end{eqnarray}
whereas conjugate momenta $p_{\phi^K}$ corresponding to the scalar field coordinates $\phi^K$ are
\begin{eqnarray}
p_{\phi^K} 
&=& 
\frac{\delta \mathcal{L}}{\delta \dot{\phi}^K} 
= 
a^3  G_{KI} \dot{\phi}^I.
\label{eq:momenta_phi}
\end{eqnarray}
Now, using Legendre transformation and the inverse transformation, the symmetry-reduced Hamiltonian $\mathcal{H}$ for the multifield scenario is immediately  obtained as
\begin{eqnarray}
\label{eq:multifield_Hamiltonian}
\mathcal{H} &\equiv& p_a \dot{a} + p_{\phi^I} \dot{\phi^I} - \mathcal{L}, \nonumber\\
&=& 
-\frac{\kappa\; {p_{a}}^2}{12 a} + \frac{G^{LM}}{2 a^3} p_{\phi^L} p_{\phi^M} + a^3 V (\phi^K),  
\end{eqnarray}
where we have used Eqs.~\eqref{eq:momenta_a} and~\eqref{eq:momenta_phi}. 

The Hamilton's equations of motion can be expressed using the Poisson bracket formalism. The symplectic structure for the geometric and matter degrees of freedom are
\begin{subequations}
\begin{eqnarray}
\label{eq:symplectic_structure_a}
\{a, p_a\}&=& 1,\\
\label{eq:symplectic_structure_phi}
\{\phi^K, p_{\phi^K}\}&=&1.
\end{eqnarray}
\end{subequations}
With this information, the equation of motion for any phase space function, $\xi$, can be elegantly written as the Poisson flow of the quantity with respect to the Hamiltonian, as follows:
\begin{align}
\label{eq:poissonbracket}
\dot{\xi} 
&= 
\{\xi, \mathcal{H}\} \nonumber\\
&= 
\biggl( 
\frac{\partial \xi}{\partial a} \frac{\partial \mathcal{H}}{\partial p_a}
- \frac{\partial \xi}{\partial p_a} \frac{\partial \mathcal{H}}{\partial a}
\biggr) + \biggl(
\frac{\partial \xi}{\partial \phi^K} \frac{\partial \mathcal{H}}{\partial p_{\phi^K}}
- \frac{\partial \xi}{\partial p_{\phi^K}} \frac{\partial \mathcal{H}}{\partial \phi^K}
\biggr).\nonumber\\    
\end{align}
Thus, the generic equations of motion for the phase space of a multifield scenario in FLRW background are
\begin{subequations}
\begin{eqnarray}
\label{eq:mutifield_a}
\dot{a} &=& -\frac{\kappa p_a}{6a},\\
\label{eq:mutifield_pa}
\dot{p}_a &=& -\frac{\kappa\; {p_a}^2}{12 a^2} + \frac{3}{2a^4} G^{IJ} p_{\phi^I} p_{\phi^J} - 3 a^2 V(\phi^K),\\
\label{eq:mutifield_phi}
\dot{\phi}^K &=& \frac{1}{a^3} G^{IK} p_{\phi^I},\\
\label{eq:mutifield_pphi}
\dot{p}_{\phi^K} &=& -a^3  V_{,K}(\phi^K) +\frac{1}{2 a^3}G^{IP}G^{JQ}p_{\phi^I} p_{\phi^J} \frac{\partial G_{PQ}}{\partial \phi^K}.
\nonumber \\
\end{eqnarray}
\end{subequations}
Evaluating the Poisson equations and using the Hamilton constraint, $\mathcal{H}=0$, leads us to the Friedmann equations and the Klein-Gordon equations, consistent with those obtained via the Euler–Lagrange formalism. The Euler–Lagrange equations for this setup have been derived in the single-field case, which we extend here to the multifield scenario. The equivalence of the two formulations, once again verifies the internal consistency.

The phase space under consideration is obtained through the symmetry reduction of the geometrodynamic variables $(q_{ab}, p^{ab})$~\cite{Ashtekar:2011ni}. Since the spatial manifold in the flat FLRW model is noncompact, a fiducial flat metric ${}^o q_{ab}$ and a fiducial cell $\mathcal{V}$ of coordinate volume $V_o$ are introduced to regulate integrals~\cite{Ashtekar:2006wn}. The physical spatial metric is $q_{ab} = a^2 \, {}^o q_{ab}$, so that physical areas and lengths are obtained by scaling fiducial ones with appropriate powers of the scale factor~\cite{Zhang:2016twe, Ashtekar:2006wn, Bojowald:1999tr}.

To capture essential quantum geometric features in cosmology, one must construct symmetry-reduced models directly at the level of the connection-dynamics phase space, rather than in the original geometrodynamic setting~\cite{Zhang:2016twe, Corichi:2008zb}.  In the isotropic case, the triad variable is related to the scale factor via $|p| = a^2 V_o^{2/3}$ and $c=\gamma \dot{a} V_o^{1/3}$.
Thus, for $V_o=1$ the canonical pair $\{c,p\}$ is related to the scale factor $a$ through~\cite{Zhang:2016twe,Wilson-Ewing:2011gnq,Corichi:2008zb}, as follows:
\begin{eqnarray}
p= a^2, \quad c= \gamma \dot{a},\nonumber    
\end{eqnarray}
where $c$ is the connection variable conjugate to the triad variable $p$, with $\gamma$ denoting the Barbero–Immirzi parameter, commonly fixed as $\gamma=0.2375$~\cite{Meissner:2004ju, Vyas:2022etz}. 
Thus, we transform the classical FLRW geometric phase space into a canonically equivalent connection dynamic phase space $\{c, p, \phi_K, p_{\phi_K}\}$, such that this transformation preserves the symplectic structure. 
In these variables, $\{c,p\}= {\kappa \gamma}/{3} $~\cite{Ashtekar:2006uz, Zhang:2016twe} and the Hamiltonian constraint takes the form
\begin{eqnarray}
\label{eq:Hamiltonian_c-p}
\mathcal{H}
= 
-\frac{3 {|p|}^{1/2} c^2}{\kappa \gamma^2}
+ 
\frac{1}{2 {|p|}^{\frac{3}{2}}} G^{IJ} p_{\phi^I} p_{\phi^J} + {|p|}^{\frac{3}{2}} V(\phi^K),
\nonumber \\
\end{eqnarray}
which is the standard starting point for LQC in the single-field case and here extended to the multifield.

Using the Poisson equation~\eqref{eq:poissonbracket}, we obtain the multifield dynamics in the $c-p$ formulation, as follows:
\begin{subequations}
\label{eq:dynamics_in_c-p}
\begin{eqnarray}
\label{eq:evolution_c}
\dot{c} &=&
-\frac{c^2}{2 \gamma |p|^{1/2}} 
-\frac{\kappa \gamma}{4 |p|^{5/2}} G_{IJ} p_{\phi^I} p_{\phi^J}    +\frac{\kappa \gamma}{2} |p|^{1/2} V ,\\ 
\label{eq:evolution_p}
\dot{p} &=& \frac{2 |p|^{1/2} c}{\gamma},\\
\label{eq:evolution_phi}
\dot{\phi^I} &=& \frac{1}{|p|^{3/2}} G^{IJ} p_{\phi^J},\\
\label{eq:evolution_pphi}
\dot{p}_{\phi^I} &=& -|p|^{3/2}  V_{,I}+\frac{1}{2 |p|^{3/2}}G^{KP}G^{JQ}p_{\phi^K} p_{\phi^J} \frac{\partial G_{PQ}}{\partial \phi^I}.
\,\,\,\,\,\,\,\,\,\,\,\,\,\,\,\,\,\,\,
\end{eqnarray}
\end{subequations}
Thus, the complete connection dynamics for a generic multiscalar field matter source with minimal coupling to gravity for FLRW spacetime are contained in Eqs.~\eqref{eq:evolution_c}--\eqref{eq:evolution_pphi}. 

The general theory of relativity does not admit a global notion of time, which is distinctly apparent in the canonical structure of the theory \cite{Barca:2021qdn, Ashtekar:2006wn, Li:2023dwy}. However, the requirement of an ``agreed clock'' demands that we consider the system in a relational way by treating one of the fields as the clock field, provided it satisfies the condition of monotonicity. Having said that, in the case of LQC, a massless scalar field serves this purpose, which later drives inflation upon the addition of a potential term at the effective level. The situation, however, is distinct in our case: we have a nontrivial coupling of the scalar fields with the kinetic term specified by the field space metric $G_{IJ}$, making the energy density nonseparable in terms of kinetic and potential contributions. This forbids switching off the potential term at the time of quantization. Since we want to construct a setup that is generic enough to allow all possibilities of interaction via $G_{IJ}$, we indeed require the introduction of an additional massless scalar field, $\Phi$, to act as a clock field. This situation has already been encountered in Ref.~\cite{Sharma:2023abm}.

\subsection{Quantum-corrected dynamics via polymerization}
\label{subsec:Quantumdynamics}

In the FLRW background, the effective dynamics are obtained via the $\bar{\mu}$ polymerization scheme (improved dynamics)~\cite{Ashtekar:2006uz, Gan:2022oiy, Ashtekar:2006wn, Zhang:2016twe} through the replacement
\begin{eqnarray}
c \;\longrightarrow\; \frac{\sin(\bar{\mu} c)}{\bar{\mu}},
\nonumber
\end{eqnarray}
where \(\bar{\mu}\) encodes the edge length of the holonomy loop and is known as polymerization parameter~\cite{Sharma:2023abm}. The holonomy loop encloses the minimum nonzero eigenvalue of area, namely the area gap \(\Delta\) such that $\bar{\mu} = \sqrt{\frac{\Delta}{|p|}}$~\cite{Ashtekar:2006wn} and \(\Delta = 4\sqrt{3}\pi \gamma \, \ell_{\text{Pl}}^2\), where $\ell_{\text{Pl}} \equiv 1/m_{\rm Pl}$
is the Planck length.

\subsection{Background dynamics}
\label{subsec:backgrounddynamics}

We thus obtain the effective Friedmann, Raychaudhuri and Klein-Gordon equations, respectively, as given by
\begin{subequations}\label{eq:Dynamics_eff}
\begin{eqnarray}
&&H^2 = \frac{\kappa \rho}{3} \left(1- \frac{\rho}{\rho_c}\right), \label{eq:FDeff}\\
&&\frac{\ddot{a}}{a} = 
-\frac{\kappa \rho}{6} \left(1 - 4 \frac{\rho}{\rho_c}\right) - \frac{\kappa P}{2}  \left(1 - 2 \frac{\rho}{\rho_c}\right) \label{eq:RCeff},\\
&&D_t \dot{\phi}^K + 3 H \dot{\phi^K} + G^{IK} V{,I} = 0. \label{eq:KGeff}
\end{eqnarray}
\end{subequations}
For later purposes, the Raychaudhari equation~\eqref{eq:RCeff} can be rephrased as,
\begin{equation}\label{eq:Hdotrephrased} 
\dot{H}(t)= -\frac{\kappa}{2} (1 + \omega) \rho \left( 1- 2\frac{\rho}{\rho_c} \right),
\end{equation}
where $\omega\equiv P/\rho$ is the equation of state.

Quantum geometry effects in LQC modify the effective Friedmann equation, Eq.~\eqref{eq:FDeff}, such that the total energy density $\rho$ attains a universal upper bound, the critical density $\rho_c$ at the bounce, given by $\rho_c = 3/(\kappa \gamma^2 \Delta)$~\cite{Ashtekar:2006uz, Ashtekar:2006wn}. At this point, the Hubble parameter vanishes, implying that the scale factor reaches a nonzero minimum and the cosmic contraction halts before transitioning into an expanding phase~\cite{Ashtekar:2006rx, Bojowald:2001xe}.

The upper bound on the energy density arises strictly from holonomy corrections to the gravitational Hamiltonian and depends only on the total energy density, independent of the specific matter content~\cite{Ashtekar:2006uz, Singh:2009mz, Ashtekar:2011ni}. 
Thus, even in multifield scenarios, $\rho$ is strictly bounded by $\rho_c$. 
From the energy density relation, Eq.~\eqref{eq:multifield_rho}, the critical value $\rho_c\approx 0.41 m_{\rm Pl}^4$ constrains the admissible values of the collective multifield velocity $\dot{\sigma}$. 

Finally, from the definition $\dot{\sigma^2} = G_{IJ} \dot{\phi}^I \dot{\phi}^J$,
we obtain the effective Klein-Gordon equation for the adiabatic field $\sigma$ as
\begin{equation}
        \ddot{\sigma} +3 H(t)\dot{\sigma} + V_{,\sigma}=0. \label{eq:resultant_KG}
\end{equation}
Upon deriving the quantum‐corrected equations, it becomes evident that the Klein–Gordon equation retains its classical form and does not receive any direct modifications from the holonomy corrections. The effects of quantum geometry enter only indirectly through the Hubble parameter $H(t)$ in the friction term, thereby influencing the field dynamics via the modified background evolution, as is the standard case in LQC in the single‐field scenario~\cite{Ashtekar:2011ni,Agullo:2023rqq,Li:2023dwy}.

Since the Eqs.~(\ref{eq:Dynamics_eff}) are not independent, it suffices to solve the Klein–Gordon equations~(\ref{eq:KGeff}) and the modified Friedmann equation~(\ref{eq:FDeff}), as the Raychaudhuri equation~(\ref{eq:RCeff}) follows from them and is thus redundant. We will focus on the two-field case such that $\phi^I= (\phi, \chi, \Phi)$ with an additional decoupled and globally monotonic relational clock field $\Phi$. 

Solving the autonomous system of coupled first- and second-order nonlinear equations yields the evolution of the dynamical variables \((a, \phi, \chi, \Phi, \dot{a}, \dot{\phi}, \dot{\chi}, \dot{\Phi})\), where the overdot denotes the derivative with respect to cosmic time. 
Once these are specified, the system is completely determined and all other physically relevant quantities and cosmological parameters such as the energy density \(\rho\), pressure density \(P\), and fractional energy density \(\Omega \equiv \rho/\rho_c\) are constructed from them. 
In this framework, the scalar fields \(\phi\) and \(\chi\) drive the dynamics, while the scale factor \(a\) governs the cosmic expansion.
To solve such a system numerically, an initial reference point in time is required, one that is necessarily traversed by this evolution. One physically motivated choice is the moment of the quantum bounce~\cite{Sharma:2018vnv} (see, however, for example Refs.~\cite{Linsefors:2013cd,Barboza:2022hng,Barboza:2020jux} and references therein for another physically motivated choice of taking initial conditions deep in the contracting phase). Since the scale factor can be freely rescaled without altering physical predictions, we fix $a=1$ at the bounce without loss of generality, i.e., $a_{\rm B}\equiv a(t_{\rm B})=1$.

In the kinetic energy dominated (KED) bounce scenario, the Universe typically evolves through three distinct phases: \emph{the bounce, transition, and slow roll}~\cite{Ashtekar:2009mm, Singh:2006im, Zhang:2007bi,  Zhu:2016dkn, Zhu:2017jew,Saeed:2024xhk,Li:2021mop}. The evolution in this case displays a universal behavior that is largely insensitive to the specific choice of initial conditions and inflationary potentials, making the KED setup  particularly advantageous for studying the preinflationary dynamics in multifield models. In contrast, potential energy dominated (PED) bounces can produce the desired slow roll only under finely tuned initial configurations. Additionally, such cases generally lack the universal postbounce dynamics characteristic of KED evolution. For a detailed study and comparison of the general features of preinflationary and inflationary phases in KED and PED bounces for a variety of single field models, we refer the reader to Refs.~\cite{Zhu:2017jew, Li:2018fco, Bhardwaj:2018omt}. In this contribution, we restrict our attention to the KED case.

To explore the sensitivity of the system to different initial conditions, we consider distinct initial values for one of the scalar fields, while that of the other fields are held the same for numerical analysis pertaining to the evolution of trajectories in both of the models we will be analyzing. Additionally, variations in the initial value of the clock field $\Phi$ have little dynamical effect on the evolution equations and, as confirmed numerically, do not significantly influence the dynamics of the system, being equivalent to a stiff fluid~\cite{Sharma:2023abm}.

To analyze the behavior of dynamical variables and related quantities in the postbounce inflationary phase, a few key parameters prove to be particularly useful. The evolution in the dominance of kinetic and potential energy largely governs these phases, making it essential to track their behavior through relevant indicators such as the equation of state parameter, \(\omega \equiv P/\rho\), which represents the effective equation of state of the cosmic fluid represented by the multiple scalar field system driving the dynamics.
It helps in identifying the distinct epochs within the preinflationary and inflationary phases.

During superinflation, where 
\begin{equation}\label{eq:superinflation_condition} 
    \dot{H}>0,
\end{equation}
the equation of state satisfies $\omega>-1$, indicating a departure from standard slow roll conditions in LQC. Conversely, during the slow roll inflationary phase, which is PED, the equation of state approaches $\omega \approx -1$. 

In addition, the two slow roll parameters are  defined as follows: the first Hubble slow roll parameter, $\epsilon_H \equiv -\dot{H}/H^2$, measures the fractional rate of change of the Hubble parameter. 
The second multifield Hubble slow roll parameter, $\eta_H \equiv \; -{\ddot{\sigma}}/{(H \dot{\sigma})}$, measures whether the accelerated expansion can be sustained for sufficiently long periods of time~\cite{Baumann:2009ds}. 
In the slow roll regime, both parameters satisfy $\epsilon_H, \eta_H \ll 1$. These conditions ensure that the Hubble rate remains nearly constant during slow roll inflation, leading to a prolonged, quasiexponential expansion. Subsequently, the Hubble rate decreases more rapidly. 

In LQC, the Friedmann equations~\eqref{eq:Dynamics_eff} acquire quantum corrections due to which the behavior of $\epsilon_{H}$ and $\eta_H$ can deviate from the standard predictions (i.e., in the usual GR case) and may lead to departures from the usual inflationary dynamics. In this context, it becomes important to examine the likelihood of occurrence of slow roll inflation as well as generation of sufficient number of e-folds during this phase. 
The total number of e-folds, $N_{\rm inf}$, is defined as usual by
\begin{equation}
\label{eq:efolds_def}
N_{\rm inf} \;=\; \ln\!\left(\frac{a_{\text{f}}}{a_{\text{i}}}\right) 
\;=\; \int_{t_{\text{i}}}^{t_{\text{f}}} H\, dt, 
\end{equation}
 which quantifies the total exponential growth of the scale factor during the inflationary epoch. Here, subscript ${\rm i}$ refers to the beginning of the inflationary regime and ${\rm f}$ refers to end of inflation.

In the following, we present the two multifield models that we will be analyzing in detail in this work.

\section{Models}\label{models}

We will present two well-motivated multifield models:

\subsection{Hybrid inflation}\label{subsec:hybridinflation}

\begin{figure}[!htb!]
  \centering
\includegraphics[scale= 0.8]{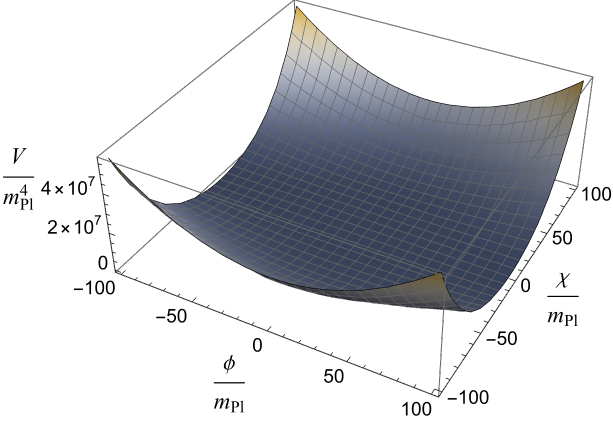}
\caption{\justifying Potential profile of the hybrid potential model Eq.~(\ref{hybridV}). This profile is obtained for the parameter values $M=4.97\times 10^{-3} \;m_{\rm Pl}, \;\lambda=\;1,\;g=\;0.8,\;m=\;0.15\; m_{\rm Pl}$.}
   \label{fig1}
\end{figure}

In hybrid inflation the potential is of the form~\cite{Linde:1993cn}
\begin{equation}
V(\phi,\chi)= \frac{\lambda}{4}(\chi^2-M^2)^2 + \frac{1}{2} m^2 \phi^2 + \frac{1}{2} g^2 \phi^2 \chi^2,
\label{hybridV}
\end{equation}
where $\phi$ denotes the inflaton and $\chi$ is known as the waterfall field. An example of the shape of the potential~(\ref{hybridV}) is
shown in {}Fig.~\ref{fig1}. Inflation proceeds
as far as $\phi> \sqrt{\lambda} M/g$. 
The action is
\begin{widetext}
\begin{eqnarray}\label{eq:Action_hybridmodel}
\!\!\!\!\!\!\!       S = \int d^4x \, \sqrt{-g} \left[ \frac{R}{2\kappa} 
- \frac{1}{2} g^{\mu \nu} \partial_\mu \phi \, \partial_\nu \phi 
- \frac{1}{2} g^{\mu \nu} \partial_\mu \chi \, \partial_\nu \chi - \frac{1}{2} g^{\mu \nu} \partial_\mu \Phi \, \partial_\nu \Phi - \frac{\lambda}{4} (\chi^2-M^2)^2
- \frac{1}{2} m^2 \phi^2 
- \frac{1}{2} g^2 \phi^2 \chi^2 
\right].
\end{eqnarray}    
\end{widetext}

Here, the field space metric $G_{IJ}$ with $\phi_1=\phi$, $\phi_2=\chi$ and $\phi_3=\Phi$ is given by the trivial metric:
\begin{eqnarray}
G_{IJ}= \begin{pmatrix}
1 & 0 & 0\\
0 & 1 & 0\\
0 & 0 & 1 \label{eq:metric_hybridmodel}
\end{pmatrix}.
\end{eqnarray}

The quantum-corrected Friedmann equations are given in Eqs.~\eqref{eq:Dynamics_eff} and \eqref{eq:Hdotrephrased}. The field evolution equations obtained from Eq.~\eqref{eq:KGeff} and corresponding to all the component fields for this model are
\begin{eqnarray}
&&\ddot{\phi} + 3H \dot{\phi}  + m^2 \phi + g^2 \chi^2 \phi = 0, 
\label{eqphi}\\
&&\ddot{\chi} + 3H \dot{\chi} + g^2 \phi^2 \chi + \lambda \chi (\chi^2 -M^2)= 0, 
\label{eqchi} \\
&&\ddot{\Phi} + 3H \dot{\Phi} = 0. 
\label{eq:KGhybridinflation}
\end{eqnarray}

The hybrid potential naturally exhibits a symmetry–breaking structure. This can be seen from the effective mass squared of the field $\chi$, which from Eq.~(\ref{hybridV}), is
given by
\begin{equation}\label{mchi2}
m_{\chi,\text{eff}}^2(\phi,\chi)
    =
    \frac{\partial^2 V}{\partial \chi^2}
    =
    g^2\phi^2
    +
    \lambda(3\chi^2 - M^2).
\end{equation}
Inflation happens when the inflaton field $\phi$ evolves along the symmetric valley with $\chi=0$, for values of $\phi>\phi_c$, where
\begin{equation}\label{eq:criticalphi}
\phi_c = \frac{\sqrt{\lambda}}{g} M.
\end{equation} 
While $\phi>\phi_c$, the discrete symmetry in the $\chi$ (the waterfall) field direction remains unbroken and the potential energy is approximately constant, dominated by the vacuum energy of the $\chi$ sector. As the inflaton rolls down its potential during the inflationary phase, the value of $\phi$ decreases. When the inflaton field drops below the critical value $\phi_c$,
the effective mass for $\chi$ becomes tachyonic and the $\chi$ field undergoes a rapid \emph{waterfall transition} towards one of the degenerate symmetry–broken vacua $\chi = \pm M$. The vacuum structure of the theory is, therefore, determined by the minima of the potential along the $\chi$ direction.  This mechanism provides a natural dynamical end to inflation and determines the vacuum structure of the model.

\subsection{String-inspired model}
\label{subsec:quartic_model}

\begin{figure}[htb!]
   \centering
\centerline{\includegraphics[scale= 0.8]{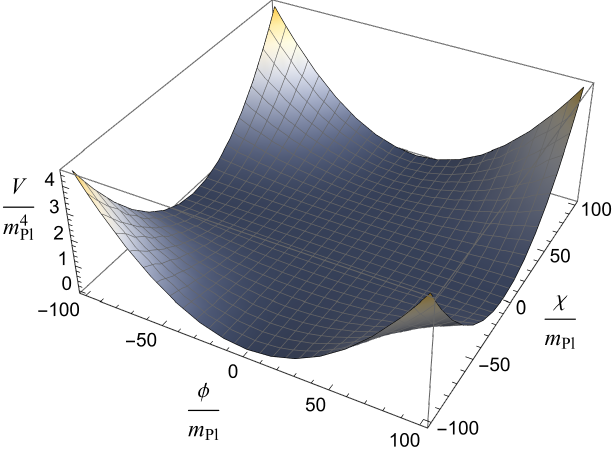}}
   \caption{\justifying Potential profile of the string-inspired model Eq.~(\ref{stringV}). The assigned values of the parameters are $m=10^{-6}{m_{\rm Pl}}$ and $g=3\times10^{-4}$.}
   \label{fig2}
\end{figure}

The second model we will analyze is characterized by the potential
\begin{equation}
V(\phi,\chi)= \frac{1}{2} m^2 \phi^2 + \frac{1}{2} g^2 \phi^2 \chi^2,
\label{stringV}
\end{equation}
with the action
\begin{widetext}
\begin{eqnarray}\label{eq:Action_quarticmodel}
       S = \int d^4x \, \sqrt{-g} \left[ \frac{R}{2\kappa} 
- \frac{1}{2} g^{\mu \nu} \partial_\mu \phi \, \partial_\nu \phi 
- \frac{e^{2b(\phi)}}{2} g^{\mu \nu} \partial _\mu \chi \, \partial_\nu \chi - \frac{1}{2} g^{\mu \nu} \partial_\mu \Phi \, \partial_\nu \Phi 
- \frac{1}{2} m^2 \phi^2 
- \frac{1}{2} g^2 \phi^2 \chi^2 
\right].
\end{eqnarray}    
\end{widetext}
An example of the shape of the potential~(\ref{stringV}) is
shown in {}Fig.~\ref{fig2}.
Note that there is an explicit kinetic coupling of $\phi$ with $\chi$. The scalar field $\phi$ here makes the role
similar to the dilaton in string-inspired models
(e.g., such as in heterotic string models~\cite{Berera:2025jbj}).
{}For this model, the field space metric $G_{IJ}$, with $\phi_1=\phi$, $\phi_2=\chi$ and $\phi_3=\Phi$, is given by
\begin{eqnarray}
G_{IJ}= \begin{pmatrix}
1 & 0 & 0\\
0 & e^{2b(\phi)} & 0\\
0 & 0 & 1 \label{eq:metric_quarticmodel}
\end{pmatrix}.
\end{eqnarray}

The kinetically coupled multifield action, characterized by a hyperbolic field-space structure~\cite{DeAngelis:2023fdu} as indicated by the scalar field metric $ G_{IJ}$, represents the second model we will analyze here, the string-inspired model. 
In this framework, the field \( \phi_1 \equiv \phi \) acts as the inflaton, decaying into the field \( \phi_2 \equiv \chi \), and $\phi_3 \equiv\Phi$ serves as a clock.

{}For the field-space metric~\eqref{eq:metric_quarticmodel}, we choose $b(\phi)= \beta \phi$, with $\beta$ an arbitrary constant~\cite{DeAngelis:2023fdu, vandeBruck:2014ata}.
{}For this model, the field evolution equations are
\begin{subequations}
\label{eq:KG_quarticmodel}
\begin{eqnarray}
\label{eq:KG_quartic_phi}
&&\ddot{\phi} +3 H(t)\dot{\phi}-\beta e^{2\beta\phi}\dot{\chi}^2+m^2\phi+g^2\chi^2\phi=0,\\
\label{eq:KG_quartic_chi}
&&\ddot{\chi} +3 H(t)\dot{\chi}+2\beta \dot{\phi}\dot{\chi}+e^{-2\beta\phi}g^2\phi^2\chi=0,\\
\label{eq:KG_quartic_Phi}
&&\ddot{\Phi} +3 H(t)\dot{\Phi}=0.
\end{eqnarray}
\end{subequations}

The quantum-corrected Friedmann equation can again be referred back to Eq.~\eqref{eq:FDeff}.
In the next two sections, we present the detailed analysis for each one of the two models specified above.

\section{Background Dynamics for the Hybrid Inflation}\label{sec:bg_hm}

\subsection{Numerical analysis}\label{subsec:hybridmodel_numericalanalysis}
   
\begin{figure}[htb!]
 \centerline{     \includegraphics[width=0.43\textwidth]{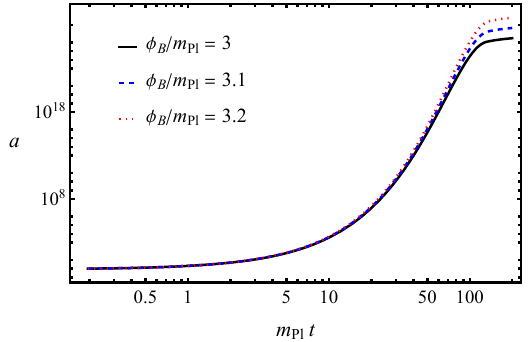}}
    \caption{\justifying Evolution of the scale factor $a(t)$ 
    for the hybrid potential model in the LQC framework. 
    The initial conditions are ($\chi_{\rm}$, $\dot{\chi}_{\rm B}$)=($0.01~m_{\rm Pl}$, $0.5~m_{\rm Pl}^2$) and $(\Phi_{\rm}, \dot{\Phi}_{\rm B})=(0.5~m_{\rm Pl}, 0.01~m_{\rm Pl}^2)$, and the set of initial values is $\phi_{\rm B}=(3.0,\,3.1,\,3.2)~m_{\rm Pl}$, where $\dot{\phi}_{\rm B}$ for each of them is determined by $\rho=\rho_c$  and $\dot{\phi}_{\rm B}>0$.
    The model parameters are chosen as $m=0.15\; m_{\rm Pl},\; g= 0.8,\; M= 4.97\times 10^{-3} \;m_{\rm Pl},\;\lambda= 1$. }
          \label{fig3}
    \end{figure}

\begin{figure}[htb!]    
\centerline{
      \includegraphics[width=0.43\textwidth]{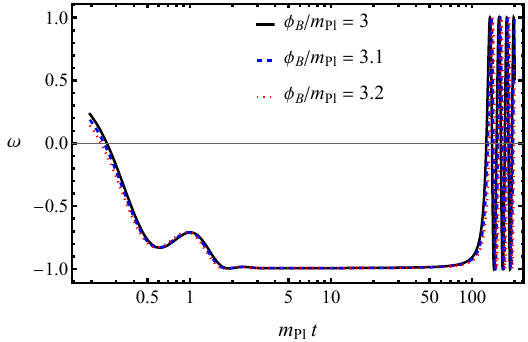}%
    }
    \caption{\justifying Evolution of the equation-of-state parameter $\omega(t)$ 
    for the hybrid potential model in the LQC framework. A smooth transition to $\omega \approx -1$ is observed. The slow roll phase concludes as $\omega$ rises above $-\frac{1}{3}$.
    The initial conditions and model parameters are the same as specified in {}Fig.~\ref{fig3}.}
    \label{fig4}
    \end{figure}

\textcolor{black}{
The evolution for the scale factor in the case of the hybrid potential of Sec.~\ref{subsec:hybridinflation} is illustrated in Fig.~\ref{fig3} for some representative initial conditions and model parameters, which will remain fixed unless we state otherwise. 
As shown in {}Fig.~\ref{fig3},  the scale factor starts to grow
exponentially for $t \gtrsim 0.3/m_{\rm Pl}$, corresponding to the start of the inflationary regime,  up to around $t \sim 100/m_{\rm Pl}$, where inflation ends.The bounce and the beginning and end of inflation, labeled from now on by the subindexes \textrm{B}, \textrm{i} and \textrm{f}, respectively, are given in Table~\ref{tab:lm_epochs} in terms of the number of e-folds $N$. Note that the larger is
$\phi_{\rm B}$, the larger is the duration of the inflationary regime.
This result also agrees with the one for the equation of state $\omega=P/\rho$ shown in {}Fig.~\ref{fig4}, where we can identify both
the beginning of inflation ($\omega < -1/3$) and its end ($\omega>-1/3$). At the end of inflation, the matter fields oscillate around their minima, with an average equation of state $\omega =0$.
}  

The quantum bounce is immediately followed by superinflation, which is defined by the condition~\eqref{eq:superinflation_condition}, corresponding to a rapid increase in the Hubble parameter. 
{}From the modified Friedmann equation in LQC, Eq.~\eqref{eq:Hdotrephrased}, since the kinetic term $( \rho +P)$ remains positive, superinflation requires the quantum correction factor $(1-2\rho/\rho_c)$ to be negative, which leads to the conditions $\rho > \rho_c/2$, $\omega >-1$, which delineate the superinflationary phase.
At the end of superinflation, $\dot{H} \to 0$, the transition phase begins and the inflaton loses kinetic energy and pressure begins to decrease, signaling the onset of a potential-dominated damping region.

    \begin{table}[tb]
    \caption{\justifying Characteristic epochs (in units of $m_{\rm Pl}^{-1}$) of the quantum-corrected background evolution for the hybrid potential model in the LQC framework in terms of $\rho$ and $\omega$, and number of e-folds of preinflation $N_{\rm preinf}$ and those of inflation $N_{\rm inf}$. The initial conditions and model parameters are the same as those considered in {}Fig.~\ref{fig3}. }
    \centering
    \renewcommand{\arraystretch}{1.2}
    \begin{tabular}{lcccc}
    \hline\hline
    &  $\phi_{\rm}/m_{\rm Pl} = $ &$3.0$& $ 3.1$ & $ 3.2$ \\
    \hline
    End of superinflation & $\rho = 0.205\,~m_{\rm Pl}^4$ & 0.69 & 0.76 & 0.81 \\
    KE$=$PE & $\omega = 0$ & 0.258 & 0.246 & 0.234 \\
    Onset of slow roll & $\omega \leq -1/3$ & 0.338 & 0.326 & 0.317 \\
    End of slow roll & $\omega \geq -1/3$ & 122.5 & 125.8 & 128.9 \\[6pt]
    \textbf{$N_{\rm preinf}=\ln(a_{\rm i}/a_{\rm B})$} & ... & 0.135 & 0.123 & 0.114 \\
    \textbf{$N_{\rm inf}=\ln(a_{\rm f}/a_{\rm i})$} & ... & \textbf{59.421} & \textbf{62.177} & \textbf{64.92} \\
    \hline\hline
    \end{tabular}
    \label{tab:lm_epochs}
    \end{table}
    
In this LQC numerical evolution, the classical regime is approached once the dynamics are sufficiently far from the bounce; that is, when the total energy density has decreased to a fraction of the critical density and quantum corrections become negligible, such that the classical Friedmann relation holds to a good approximation.
    
\begin{center}
\begin{figure}[!htb]
\subfigure[]{\includegraphics[width=7.5cm]{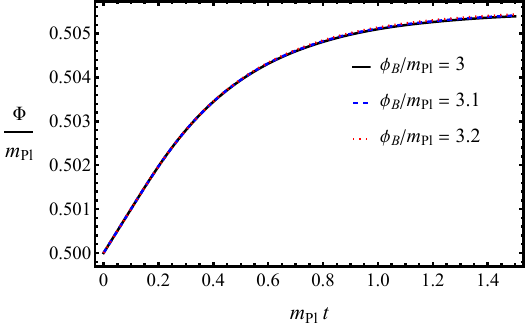}}
\subfigure[]{\includegraphics[width=7.5cm]{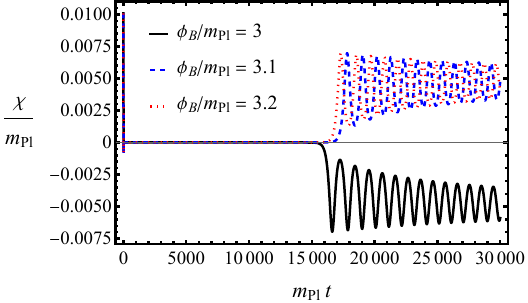}}
\subfigure[]{\includegraphics[width=7.5cm]{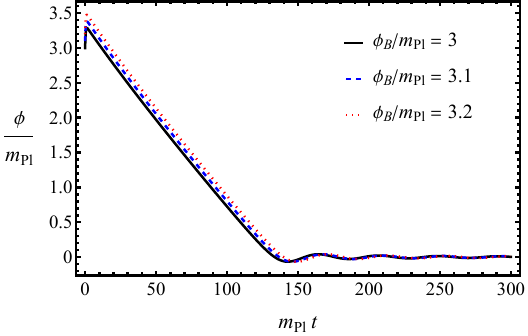}}
\caption{\justifying Evolution for each one of the fields, $\Phi(t)$ (panel a), $\chi(t)$ (panel b), and $\phi(t)$ (panel c), for the hybrid potential model in the LQC framework. The initial conditions and model parameters are the same ones considered in {}Fig.~\ref{fig3}.}
  \label{fig5}
\end{figure}
\end{center}

\textcolor{black}{
The dynamical behavior for each field in the model, starting from right after the bounce, $t\geq t_{\rm B}$, up to after the inflationary regime, $t>t_{\rm f}$, is shown in
{}Fig.~\ref{fig5}. The model parameters and initial conditions
are the same as the ones used in {}Fig.~\ref{fig3}.
The behavior of the clock field $\Phi$ is shown in {}Fig.~\ref{fig5}(a). {}From the bounce until the beginning of inflation,  $\Phi$ evolves monotonically. During inflation, it asymptotes
to a constant value that depends on the chosen initial conditions. We can see this from its evolution equation (\ref{eq:KGhybridinflation}). By considering that during slow roll $H$ can be approximated as a constant $H_i\equiv H(t_i)$, where $t_i$ is the beginning of the slow roll inflationary regime, the solution of Eq.~(\ref{eq:KGhybridinflation}) can be approximated by $\Phi(t) \sim \Phi(t_i) +\dot\Phi(t_i)/(3H_i) - \dot \Phi(t_i) e^{-3H_i (t-t_i)}/(3H_i)$.
We can then see that during inflation, the clock field tends to go asymptotically to
the constant value  $\Phi(t_i) +\dot\Phi(t_i)/(3H_i)$ for $t \gg t_i$ but with still $t< t_f$.
Likewise, the waterfall field $\chi$ goes asymptotically to its vacuum expectation value $\langle \chi \rangle = \pm M$ after the inflaton field $\phi$ goes
below the critical $\phi_c$ given by Eq.~(\ref{eq:criticalphi}). While $m_\chi = g \phi < 3H$, the dynamics for $\chi$ is overdamped, with the Hubble friction keeping $\chi$ frozen at $\chi =0$. When $m_\chi > 3 H$, the dynamics of $\chi$ are underdamped and $\chi$ will then oscillate around either $+M$ or $-M$ as seen in {}Fig.~\ref{fig5}(b).
{}Finally, the inflaton field $\phi$ decreases monotonically until  the  end of
inflation, after which it oscillates around its minimum 
value $\phi =0$ as seen in {}Fig.~\ref{fig5}(c).
}

\begin{figure*}[htb!] 
\centerline{         \includegraphics[width=1\textwidth]{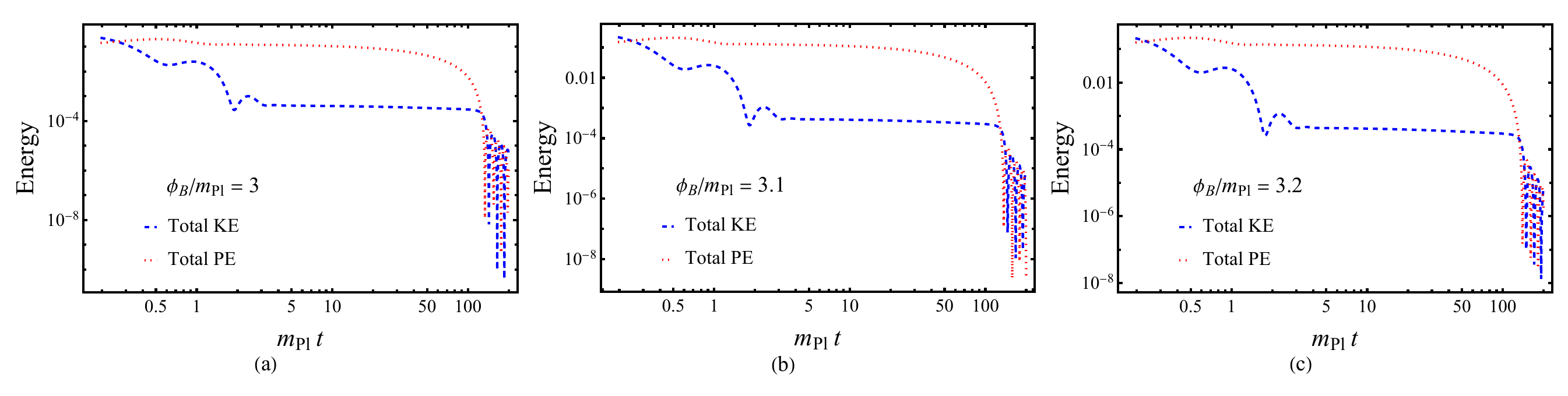}}
    \caption{\justifying Comparison of the kinetic energy and potential energy during the preinflationary and inflationary evolution of the hybrid model. The initial conditions and model parameters are the same ones considered in {}Fig.~\ref{fig3}, with
    each panel corresponding to a distinct value of $\phi_{\rm B}$.}
    \label{fig6}  
\end{figure*}

\textcolor{black}{
The behavior of the energy densities can also be understood from the evolution equations for the fields, Eqs.~(\ref{eqphi}), (\ref{eqchi}) and (\ref{eq:KGhybridinflation}), alongside with the modified {}Friedmann equation in LQC, Eq.~\eqref{eq:FDeff}. 
Starting at the bounce instant, where
$H=0$ and $\dot{H}>0$, the total energy density decreases as $\dot{\rho} \simeq -3H(\dot{\phi}^2 + \dot{\chi}^2 + \dot{\Phi}^2) < 0$. Close to the bounce the total kinetic energy $(\dot{\phi}^2 + \dot{\chi}^2 + \dot{\Phi}^2)/2$ dominates over the total potential energy. The system evolves like stiff matter close to the bounce with
$\rho_{\rm total} \propto 1/a^6$. Soon after the bounce, the potential energy (mainly set by that of the inflaton) becomes dominant and the inflationary regime begins. After inflation, 
the fields oscillate around their vacuum expectation values and the potential energy becomes again subdominant. These different  KED and PED regimes are illustrated in {}Fig.~\ref{fig6}.
}
    
Since the dynamics of the inflaton field are related to that of the waterfall field, let us analyze in some more detail the dynamics of $\chi$.
To analyze the stability of the field $\chi$, one considers its effective mass squared $m_{\chi,\text{eff}}^2(\phi)$, Eq.~\eqref{mchi2}, along the valley $\chi=0$.
The sign of $m_{\chi,\text{eff}}^2(\phi)$ determines the stability in the direction of the field $\chi$. Evaluating it for $\phi_{\rm B}= \{3.0,3.1,3.2\}~m_{\rm Pl}$, we obtain $m_{\chi,\text{eff}}^2(\phi_{\rm B}) \approx \{5.759, 6.150, 6.553\}~m_{\rm Pl}^2$, respectively, at the bounce. 
The critical value of the inflaton given by Eq.~\eqref{eq:criticalphi} at which the waterfall instability sets in is $\phi_c \approx 0.0062\,m_{\rm Pl}$.
Since the aforementioned initial inflaton values satisfy $\phi_{\rm B}\gg\phi_c$, the system begins its evolution already inside the inflationary valley. Let us explore this regime before we consider the waterfall one.
    
The true vacuum state is obtained by minimizing the potential with respect to both the scalar fields, which yields $(\phi,\chi)=(0,\pm M)=(0, \pm  \; 4.97\times 10^{-3}\;m_{\rm Pl})$, leading to $V=0$ and corresponding to the global minima of the potential. In contrast, the point $(\phi,\chi)=(0,0)$ corresponds to a local maximum along the $\chi$-field direction with $V(0,0)=\lambda M^4/4=1.525 \times 10^{-10}\,m_{\rm Pl}^4$ and therefore represents a false vacuum state.
Soon after the KED bounce, the dynamics are dominated by the kinetic energy of the scalar fields. As the Universe expands, Hubble friction rapidly dilutes these kinetic contributions (see Fig.~\ref{fig6}), while the false vacuum energy remains constant. 
Once the kinetic energy becomes subdominant, the total energy density is effectively controlled by this vacuum contribution, leading to an approximately constant Hubble rate determined by $3H^2\simeq \kappa V(0,0)(1-{V(0,0)}/{\rho_c})$, and thus primarily driving slow roll instead of being driven by the flatness of the inflaton potential alone.

Soon after the superinflation ends (see Table~\ref{tab:lm_epochs}), we have $\omega = 0$
for each value of $\phi_{\rm B}$ indicating that the kinetic and potential energies are equal.
Ultimately, irrespective of the initial values, we can observe $\omega \approx -1$ in {}Fig.~\ref{fig4}, indicating the slow roll phase.
    
During superinflation, the slow roll coefficient $\epsilon_H =-\dot H/H^2< 0$ since $\dot{H} > 0$. At the epoch given in Table~\ref{tab:lm_epochs}, $\epsilon_H$ crosses zero, marking the end of superinflation. The system then enters the damping region, after which $\epsilon_H \ll 1$ during slow roll. 
        
As the Universe enters the expanding branch ($H>0$), both fields experience Hubble friction, but their initial velocities determine the direction of motion in field space. We examine the configuration in which both the inflaton $\phi$ and $\chi$  scalar fields emerge from the quantum bounce with positive or negative initial velocities, i.e., $\dot{\phi}_{\rm B}$ and $\dot{\chi}_{\rm B}$ greater or small than zero. 
    
Quantitative exploration for the representative parameters and initial conditions $M = \,m_{\rm Pl},\ \lambda = 0.3,\ g = 10^{-5},\ m = 0.01\,m_{\rm Pl}$, with the clock field fixed at $(\Phi_{\rm B}, \dot{\Phi}_{\rm B}) = (0.5\,m_{\rm Pl},\,0.5\,m_{\rm Pl}^2)$ and $\dot{\chi}_{\rm B} = \pm 0.05\,m_{\rm Pl}^2$, yields $\phi_c \approx 5.48 \times 10^{4}\,m_{\rm Pl}$. Since the initial inflaton values satisfy $\phi_{\rm B} \ll \phi_c$, the system begins its evolution already inside the tachyonic regime.
Quantitatively sweeping $\phi_{\rm B}$ in the range $[0.001,\, 8]\,m_{\rm Pl}$ while keeping $\chi_{\rm B}$ fixed leads to a mild increase of $N_{\mathrm{inf}}$, showing that the e-fold count quickly saturates at relatively small values of $\phi_{\rm B}$ and exhibits little further enhancement thereafter. The complementary variation with respect to the auxiliary field $\chi_{\rm B}$ reveals a much stronger sensitivity: as $\chi_{\rm B}$ increases in the range [$10^{-5},\,0.4]\;m_{\rm Pl}$, $N_{\mathrm{inf}}$ drops sharply. 

In contrast, the case in which both fields begin with negative initial velocities, $\dot{\phi}_{\rm B}<0$ and $\dot{\chi}_{\rm B}<0$, exhibits significantly different behavior. 
Varying $\phi_{\rm B}$ produces large $N_{\mathrm{inf}}$ only when the initial amplitude is small. Once the inflaton amplitude grows beyond ${\sim}6\;m_{\mathrm{Pl}}$, the system shows significant decline in the total e-folds. A similar trend is observed when varying $\chi_{\rm B}$,
where even moderate increases in $\chi_{\rm B}$ sharply reduce the duration of inflation.
    
The third scenario corresponds to mixed initial velocities, with $\dot{\phi}_{\rm B}<0$ and $\dot{\chi}_{\rm B}>0$. 
Varying $\phi_{\rm B}$ in this case results in only modest changes in $N_{\mathrm{inf}}$, with the e-folds remaining relatively stable over a wide range of initial amplitudes. The variation with respect to $\chi_{\rm B}$
again produces a monotonic decrease in $N_{\mathrm{inf}}$ as $\chi_{\rm B}$ increases.
    
 In the remaining mixed case with $\dot{\phi}_{\rm B}>0$ and $\dot{\chi}_{\rm B}<0$, varying $\phi_{\rm B}$ leads to only small changes in $N_{\mathrm{inf}}$, 
 while sweeping $\chi_{\rm B}$ produces large declines in the duration of inflation.
A quantitative examination of the e-fold ranges highlights the asymmetry in sensitivity between the inflaton and the auxiliary waterfall $\chi$ field. Sweeping $\chi_{\rm B}$ across its allowed range produces large changes in $N_{\mathrm{inf}}$ in every configuration. By contrast, sweeping $\phi_{\rm B}$ changes the total e-folds by less than 1.1 in three of the four cases, indicating that $N_{\mathrm{inf}}$ saturates rapidly once $\phi_{\rm B}$ exceeds a small threshold. The single exception is the fully negative configuration, where large $\phi_{\rm B}$ drives the trajectory away from the slow roll region. In other words, $\chi_{\rm B}$ effectively sets the conducive condition for whether slow roll is achieved, whereas $\phi_{\rm B}$ mostly regulates how long inflation lasts once it is reached.
{}For these parameter values, none of the configurations yield the canonical $N_{\mathrm{inf}}\gtrsim 60$ e-folds required for observational consistency in standard cosmology. Successful long-duration inflation in the hybrid model in LQC therefore likely requires either carefully tuned initial data or choices of coupling parameters. Some successful cases are exhibited in Tables~\ref{tab:Ninf_velocity_cases_KED_pp_mm} and ~\ref{tab:Ninf_velocity_cases_KED_pm_mp}. 

\begin{table}[htb!]
\caption{\justifying Number of preinflationary $N_{\rm preinf}$ and inflationary e-folds $N_{\rm inf}$ for same velocity sign configurations at the bounce in the KED scenario with $\phi_{\rm B}>\phi_c$, $\phi_c=0.0062 m_{\rm Pl}$, using Eq.~\eqref{eq:criticalphi}. The model parameter values are $m = 1.50\times10^{-1}m_{\rm Pl}$, $g = 8\times10^{-1}$, $M = 4.97\times10^{-3}m_{\rm Pl}$, and $\lambda = 1$~\cite{Copeland:1994vg}. The initial condition for the clock field is chosen as $\Phi_{\rm B} = 0.5\,m_{\rm Pl}$ and $\dot \Phi_{\rm B} = 0.01\,m_{\rm Pl}^2$ in all cases. $\dot{\chi}_{\rm B} = -0.5\,m_{\rm Pl}^2$ when  $\dot{\chi}_{\rm B}<0$ and $\dot{\chi}_{\rm B} = 0.5\,m_{\rm Pl}^2$ when  $\dot{\chi}_{\rm B}>0$.
}
\centering
\resizebox{\columnwidth}{!}{
\begin{tabular}{
l c c c @{\hspace{10pt}}
l c c c
}
\hline\hline
\noalign{\vskip 4pt}
\multicolumn{4}{l}{$(\dot{\phi}_{\rm B},\dot{\chi}_{\rm B})=(+,+)$} &
\multicolumn{4}{c}{$(\dot{\phi}_{\rm B},\dot{\chi}_{\rm B})=(-,-)$} \\

\cmidrule(lr){1-4}\cmidrule(lr){5-8}

$\phi_{\rm B}/ m_{\rm Pl}$ & $\chi_{\rm B}/ m_{\rm Pl}$ & $N_{\rm preinf}$ & $N_{\rm inf}$ &
$\phi_{\rm B}/ m_{\rm Pl}$ & $\chi_{\rm B}/ m_{\rm Pl}$ & $N_{\rm preinf}$ & $N_{\rm inf}$ \\

\hline\\[-4pt]

2 & 0.01 & 0.252 & 33.354   
  & 3   & 0.001   & 0.230 & 42.452 \\

\textbf{3} & \textbf{0.01} & \textbf{0.137} & \textbf{59.419}
  & 3.2 & 0.0001  & 0.207 & 48.287 \\

3.1 & 0.01 & 0.127 & 62.166 
    & 3.3 & $10^{-5}$ & 0.186 & 51.269 \\

3.2 & 0.01 & 0.117 & 64.920
    & 3.4 & $10^{-5}$ & 0.174 & 54.274 \\

3.3 & 0.01 & 0.107 & 67.673 
    & \textbf{3.6} & \textbf{\boldmath$10^{-6}$} & \textbf{0.151} & \textbf{60.377} \\

\hline\hline
\end{tabular}
}
\label{tab:Ninf_velocity_cases_KED_pp_mm}
\end{table}

\begin{table}[htb!]
\caption{\justifying
 Number of preinflationary $N_{\rm preinf}$ and inflationary e-folds $N_{\rm inf}$ for opposite velocity sign configurations at the bounce for the same conditions and parameters for which Table~\ref{tab:Ninf_velocity_cases_KED_pp_mm} is evaluated. 
}
\centering
\resizebox{\columnwidth}{!}{
\begin{tabular}{
l l c c @{\hspace{10pt}}
l c c c
}
\hline\hline
\noalign{\vskip 4pt}
\multicolumn{4}{l}{$(\dot{\phi}_{\rm B},\dot{\chi}_{\rm B})=(+,-)$} &
\multicolumn{4}{c}{$(\dot{\phi}_{\rm B},\dot{\chi}_{\rm B})=(-,+)$} \\

\cmidrule(lr){1-4}\cmidrule(lr){5-8}

$\phi_{\rm B}/ m_{\rm Pl}$ & $\chi_{\rm B}/ m_{\rm Pl}$ & $N_{\rm preinf}$ & $N_{\rm inf}$ &
$\phi_{\rm B}/ m_{\rm Pl}$ & $\chi_{\rm B}/ m_{\rm Pl}$ & $N_{\rm preinf}$ & $N_{\rm inf}$ \\

\hline\\[-4pt]

3   & 0.1     & 0.208 & 58.638   
  & 3.3 & 0.01    & 0.140 & 51.300 \\

3   & 0.01    & 0.148 & 59.407
  & 3.5 & 0.001   & 0.162 & 57.305 \\

\textbf{3.05} & \textbf{0.0001} & \textbf{0.138} & \textbf{60.789} 
  & 3.5 & 0.0001 & 0.189 & 57.280 \\

3.1 & 0.001  & 0.137 & 62.157 
  & \textbf{3.6} & \textbf{0.0001} & \textbf{0.145} & \textbf{60.275} \\

3.2  & 0.01 & 0.128 & 64.890   
  & 3.7 & $10^{-5}$ & 0.177 & 63.288 \\

\hline\hline
\end{tabular}
}
\label{tab:Ninf_velocity_cases_KED_pm_mp}
\end{table}
    
{}From Tables~\ref{tab:Ninf_velocity_cases_KED_pp_mm} and ~\ref{tab:Ninf_velocity_cases_KED_pm_mp}, 
we observe that for the KED bounce, the model consistently yields a slow roll inflationary phase with approximately $60$ e-folds of inflation before the waterfall transition is reached, for suitable parameter choices following Ref.~\cite{Copeland:1994vg}. The highlighted rows indicate representative initial conditions for which $N_{\rm inf}\simeq 60$ is achieved. This demonstrates that by appropriately fine-tuning the initial phase space variables at the bounce, the model remains viable within the LQC framework  irrespective of the sign of the initial field velocities.

The number of e-folds of expansion during the preinflationary phase, across the scanned phase space and for all configurations of the initial velocity signs, is $N_{\mathrm{preinf}} \approx 0.1-0.4$. This indicates that the preinflationary regime contributes only a very small amount to the total expansion of the Universe, with the dominant growth occurring during the subsequent slow roll inflationary phase.

{}For completeness, the analytical background solutions corresponding to the bounce, superinflationary, and slow roll phases are presented in Appendix~\ref{sec:analytical_solutions}. These derivations provide explicit expressions for the field, scale factor, and Hubble evolution under the same approximations employed in the numerical analysis, allowing a direct comparison between the analytical and numerical treatments of the model. These solutions are evaluated in terms of the adiabatic field $\sigma$, rather than the individual field components, to capture the collective inflationary dynamics of the system.
    
\subsection{Qualitative analysis: Dynamical system approach}\label{subsec:hybridmodel_dynamicalanalysis}

Using the chosen configuration of potential Eq.~\eqref{hybridV} and $G_{IJ}$ metric~\eqref{eq:metric_hybridmodel}, we can express the modified Friedmann Eq.~\eqref{eq:Dynamics_eff} and evolution equations for the fields, Eqs.~(\ref{eqphi}), (\ref{eqchi}) and (\ref{eq:KGhybridinflation}), into an autonomous system of first-order differential equations, whose critical points determine the asymptotic cosmological behavior and allow a stability analysis. In order to do so, we define a set of dimensionless dynamical variables that normalize the fields' kinetic energies, components of potential energy, and the relative energy density. Thus, the definitions of dynamical variables are adapted from the literature~\cite{Cicoli:2020noz, Samart:2007xz, Das:2023rat, Dimakis:2019qfs, Sharma:2018vnv, Shahalam:2017wba} as follows:

\begin{align}\label{eq:dynvar_hybridinflation}
x_\phi &\equiv  \frac{\dot{\phi}}{\sqrt{2\rho_c}}, \quad &
x_\chi &\equiv  \frac{\dot{\chi}}{\sqrt{2\rho_c}} , \quad &  
x_\Phi &\equiv \frac{\dot{\Phi}}{\sqrt{2\rho_c}},\nonumber
\\
y_\phi &\equiv  \frac{m \phi}{\sqrt{2\rho_c}}, \quad &
y_\chi &\equiv \frac{\chi}{M}, \quad &
y_\Phi &\equiv \frac{\Phi}{M} .
\end{align}

Additionally, we define a parameter $z \equiv \frac{\rho}{\rho_c}$.
Substituting the dimensionless variables into the energy density equation and normalizing it with critical density yields a constraint that must be satisfied by all physical trajectories in the phase space, as follows:
\begin{equation}
z= x_\phi^2 + x_\chi^2 + x_\Phi^2 + y_\phi^2 + \frac{g^2 M^2}{m^2} y_\phi^2 y_\chi^2 + \frac{\lambda M^4}{4 \rho_c} (y_\chi^2-1)^2. \label{eq:constraint_hybridmodel}
\end{equation}
The modified Friedmann equation~\eqref{eq:FDeff} then becomes
\begin{equation}\label{eq:FDdimless}
    H^2 = \frac{\kappa \rho_c}{3} z (1-z).
\end{equation}

In LQC, the modified {}Friedmann equation imposes a fundamental upper bound on the energy density, as follows:
\begin{equation}\label{eq:zbound}
0 \le z \equiv \frac{\rho}{\rho_c} \le 1.
\end{equation}
Consequently, the dynamical variables cannot take arbitrary values, since they must satisfy the energy density constraint~\eqref{eq:constraint_hybridmodel}. The physically admissible phase space is therefore not the entire dynamical space, but a bounded submanifold defined by the condition~(\ref{eq:zbound}). In order to analyze the global structure of the phase space and correctly identify the allowed locus of fixed points, it is convenient to compactify this bounded variable.
To achieve this, we introduce a trigonometric parametrization~\cite{Boehmer:2014vea},
\begin{equation}
z = \sin^2\theta,
\qquad
0 \le \theta \le \pi,
\label{eq:zcompact}
\end{equation}
which automatically enforces the physical bound on the energy density. In terms of this variable, the modified Friedmann equation gives
\begin{equation}
H
=
\pm
\sqrt{\frac{\kappa\rho_c}{12}}\,
\sin(2\theta).
\label{eq:H_theta}
\end{equation}

Equation~\eqref{eq:H_theta} shows that the Hubble parameter naturally splits into two branches:
(1) the positive branch $H>0$, corresponding to the expanding Universe; and (2) the negative branch $H<0$, corresponding to the contracting Universe.
The bounce occurs at $z=1$, or equivalently $\theta=\tfrac{\pi}{2}$, where $H=0$. 
Similarly, $z=0$ corresponds to $\theta=0$ or $\theta=\pi$, representing the classical low–density regime. 
Thus, the trigonometric compactification provides a global parametrization of the physical phase space, smoothly connecting the contracting and expanding branches through the bounce.

In terms of the dynamical variables, the energy density constraint now takes the compact form
\begin{equation}
\sin^2\theta
=
x_\phi^2 + x_\chi^2 + x_\Phi^2
+ y_\phi^2
+ \frac{g^2 M^2}{m^2} y_\phi^2 y_\chi^2
+ \frac{\lambda M^4}{4 \rho_c} (y_\chi^2-1)^2.
\label{eq:constraint_theta}
\end{equation}
This representation makes the bounded nature of the physical phase space explicit and ensures that the autonomous system evolves on a compact submanifold. Fixed points must therefore satisfy not only the vanishing of the dynamical derivatives, but also the compactified constraints~\eqref{eq:H_theta} and \eqref{eq:constraint_theta}.

We introduce the dimensionless time variable $\tau = M t$ so that the time derivative transforms as $\frac{d}{d\tau} = \frac{1}{M}\frac{d}{dt}$. Using this relation and the definitions in Eq.~\eqref{eq:dynvar_hybridinflation}, the autonomous system becomes
\begin{align}
\frac{dx_\phi}{d\tau}
&=
-\frac{3H}{M}x_\phi
-\frac{m}{M}y_\phi
-\frac{g^2M}{m}y_\phi y_\chi^2,
\\[6pt]
\frac{dx_\chi}{d\tau}
&=
-\frac{3H}{M}x_\chi
-\frac{g^2 \sqrt{2 \rho_c}}{m^2}y_\phi^2 y_\chi
-\frac{\lambda M^2}{\sqrt{2\rho_c}}
y_\chi(y_\chi^2-1),
\end{align}
\begin{align}
\frac{dx_\Phi}{d\tau}
&=
-\frac{3H}{M}x_\Phi,
\end{align}
\begin{align}
\frac{dy_\phi}{d\tau}
&=
\frac{m}{M}x_\phi,
\\[6pt]
\frac{dy_\chi}{d\tau}
&=
\frac{\sqrt{2\rho_c}}{M^2}\,x_\chi,
\end{align}
\begin{align}
\frac{dy_\Phi}{d\tau}
&=
\frac{\sqrt{2\rho_c}}{M^2}\,x_\Phi.
\end{align}

\begin{table*}[!htb]
\caption{\justifying
Critical points for the hybrid inflation model in LQC dynamical system. An overbar denotes an arbitrary constant value along that direction, indicating a continuous locus of equilibrium points rather than an isolated critical point. 
Such nonisolated equilibria are nonhyperbolic and their stability cannot be fully determined from linear analysis alone, requiring center manifold or higher-order analysis~\cite{Lakshmanan:2003uaa}.}
\centering
\label{tab:fixedpoints_hybrid}

\begin{tabular}{c c c c}
\hline \hline
Point & Fixed point coordinates & Eigenvalues & Linear stability\\
\hline

\multicolumn{4}{l}{Expanding branch ($H>0$)}\\

$P_1^{(+)}$ &
$\begin{aligned}
&(x_\phi,x_\chi,x_\Phi,y_\phi,y_\chi,y_\Phi)\\
&=(0,0,0,0,-1,\bar{y}_\Phi)
\end{aligned}$
&
$\{\pm0.7746 i,\;\pm0.01 i,\;0,\;0\}$
&
Nonhyperbolic center
\\

$P_2^{(+)}$ &
$\begin{aligned}
&(x_\phi,x_\chi,x_\Phi,y_\phi,y_\chi,y_\Phi)\\
&=(0,0,0,0,0,\bar{y}_\Phi)
\end{aligned}$
&
$\{-2.2810,\,-2.1495,\,-2.1495,\,0.1315,\,-4.65\times10^{-5},\,0\}$
&
Unstable saddle
\\

$P_3^{(+)}$ &
$\begin{aligned}
&(x_\phi,x_\chi,x_\Phi,y_\phi,y_\chi,y_\Phi)\\
&=(0,0,0,0,1,\bar{y}_\Phi)
\end{aligned}$
&
$\{\pm0.7746 i,\;\pm0.01 i,\;0,\;0\}$
&
Nonhyperbolic center
\\

\noalign{\vskip 10pt}
\multicolumn{4}{l}{Contracting branch ($H<0$)}\\

$P_1^{(-)}$ &
$\begin{aligned}
&(x_\phi,x_\chi,x_\Phi,y_\phi,y_\chi,y_\Phi)\\
&=(0,0,0,0,-1,\bar{y}_\Phi)
\end{aligned}$
&
$\{\pm0.7746 i,\;\pm0.01 i,\;0,\;0\}$
&
Nonhyperbolic center
\\

$P_2^{(-)}$ &
$\begin{aligned}
&(x_\phi,x_\chi,x_\Phi,y_\phi,y_\chi,y_\Phi)\\
&=(0,0,0,0,0,\bar{y}_\Phi)
\end{aligned}$
&
$\{2.2810,\,2.1495,\,2.1495,\,-0.1315,\,4.65\times10^{-5},\,0\}$
&
\makecell{Unstable saddle\\ (time-reversed directions)}
\\

$P_3^{(-)}$ &
$\begin{aligned}
&(x_\phi,x_\chi,x_\Phi,y_\phi,y_\chi,y_\Phi)\\
&=(0,0,0,0,1,\bar{y}_\Phi)
\end{aligned}$
&
$\{\pm0.7746 i,\;\pm0.01 i,\;0,\;0\}$
&
Nonhyperbolic center
\\
\hline \hline

\end{tabular}

\end{table*}

To study the stability of this model, we linearize the system by expanding the evolution equations around the fixed points and retaining terms up to first order in perturbations. This leads to a system of linear differential equations whose coefficients are encapsulated in the Jacobian matrix given by $\mathcal{J}_{ij} \equiv \frac{\partial F_i}{\partial X_j}$ where $F_i \in 
\left\{
\frac{dx_\phi}{d\tau},
\frac{dx_\chi}{d\tau},
\frac{dx_\Phi}{d\tau},
\frac{dy_\phi}{d\tau},
\frac{dy_\chi}{d\tau},
\frac{dy_\Phi}{d\tau}
\right\}$ and $X_j \in \{x_\phi,\, x_\chi,\, x_\Phi,\, y_\phi,\, y_\chi,\, y_\Phi\}$.
The fixed points of the system are obtained by requiring the simultaneous vanishing of all autonomous derivatives,
\begin{align}\label{eq:fixedpoint_condition_hybrid}
\frac{dx_\phi}{d\tau}
=
\frac{dx_\chi}{d\tau}
=
\frac{dx_\Phi}{d\tau}
=
\frac{dy_\phi}{d\tau}
=
\frac{dy_\chi}{d\tau}
=
\frac{dy_\Phi}{d\tau}
= 0 ,
\end{align}
and then solving the resulting algebraic equations subject to the constraints~\eqref{eq:FDdimless} and~\eqref{eq:constraint_hybridmodel} as presented in Table~\ref{tab:fixedpoints_hybrid}. These points correspond to the asymptotic cosmological states of the model, and their stability analysis follows from a linearization of this autonomous system around each critical point.

The dynamical system analysis reproduces the structure given by true and false vacua discussed in Sec.~\ref{subsec:hybridinflation}, at the level of phase–space fixed points. The fixed point $P_2$ corresponds to the configuration $(\phi,\chi)=(0,0)$, where all kinetic terms vanish and the energy density is entirely dominated by the constant potential. This configuration therefore represents the false vacuum state.

The stability analysis shows that $P_2$ is a saddle point: in the expanding branch, some directions are unstable, while in the contracting branch the signs of the real parts of the eigenvalues are reversed. This behavior reflects the time–reversal symmetry of the Hubble friction term, which changes sign between the expanding and contracting phases.
The equation of state at this point is $\omega=\frac{P}{\rho}=-1$, which corresponds to a vacuum–dominated or inflationary state.

In contrast, the fixed points $P_1$ and $P_3$ correspond to the configurations
\begin{equation}
(\phi,\chi)=(0,-M),
\qquad
(\phi,\chi)=(0,+M),
\end{equation}
which are precisely the true vacuum states of the theory. At these points, all kinetic energies vanish, the potential energy is zero, and the total energy density satisfies $\rho=0$. Consequently, the Hubble parameter also vanishes, and the Universe approaches a static, zero–energy configuration. These points, therefore, represent the late–time vacuum states reached after the waterfall transition.

The eigenvalues at $P_1$ and $P_3$ are purely imaginary, with additional zero eigenvalues, and are identical in both the expanding and contracting branches. These fixed points then form nonhyperbolic centers corresponding to a continuous surface of equilibria parametrized by the $\chi$ and clock–field directions. As discussed in the standard dynamical systems literature~\cite{Lakshmanan:2003uaa}, when fixed points form a continuous set rather than isolated points, linear stability analysis becomes inconclusive along the neutral directions. In such cases, the dynamics are governed by the center manifold, and higher–order or numerical analysis is required to determine the ultimate stability properties of the system. Physically, this degeneracy arises because the clock field is free and does not contribute to the potential energy, leading to a conserved direction in phase space.

{}From a global phase–space perspective, the cosmological evolution in this model can therefore be understood as a transition from the false vacuum fixed point $P_2$ to one of the true vacuum fixed points $P_1$ or $P_3$. Inflation proceeds while the system remains close to the false vacuum configuration. Once the critical value $\phi_c$ is reached, the $\chi$ direction becomes unstable, triggering the waterfall transition. The trajectory then rapidly evolves toward one of the symmetry–broken minima, where it settles into the corresponding vacuum fixed point. This dynamical systems picture, therefore, reproduces the standard hybrid inflation scenario.

\section{Background Dynamics for the String-inspired Model}\label{sec:bg_sm}

\subsection{Numerical analysis}\label{subsec:quarticmodel_numericalanalysisnew}

\begin{figure}[htb!]
\centerline{  \includegraphics[width=0.43\textwidth]{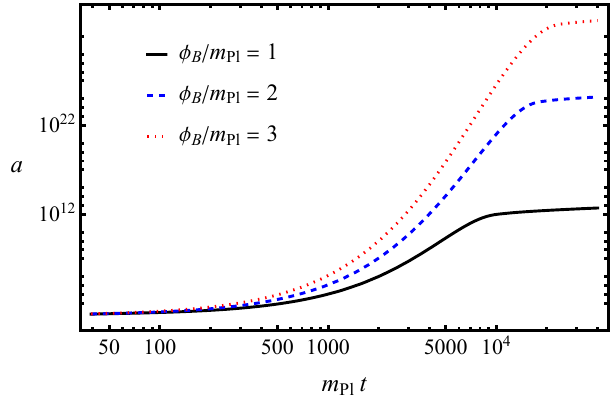}%
}
\caption{\justifying Evolution of the scale factor $a(t)$,
for the string-inspired model~(\ref{eq:Action_quarticmodel}) with a noncanonical hyperbolic kinetic space represented by the metric~(\ref{eq:metric_quarticmodel}) in the LQC framework. 
The set of initial values of the field $\phi$ are taken as 
$\phi_{\rm B}=(1,\,2,\,3)~m_{\rm Pl}$. $\dot{\phi}_{\rm B}$ is determined by $\rho=\rho_c$ and $\dot{\phi}_{\rm B}>0$ in each case.
The $\chi_{\rm B}$ value is fixed at $4.5~m_{\rm Pl}$ and its velocity at $\dot{\chi_{\rm B}}= 0.1~m_{\rm Pl}^2 $. The initial values of the clock field and its velocity are fixed at $(\Phi_{\rm B}, \dot{\Phi}_{\rm B})=(0.5~m_{\rm Pl}, 0.5~m_{\rm Pl}^2)$. The model parameters are $m=10^{-6}\; m_{\rm Pl},\; g= 3\times10^{-4},\; and \;  \beta=-0.1 m_{\rm Pl}^{-1}$.}
\label{fig7}
\end{figure}

\begin{figure}[!htb]
\centerline{
  \includegraphics[width=0.43\textwidth]{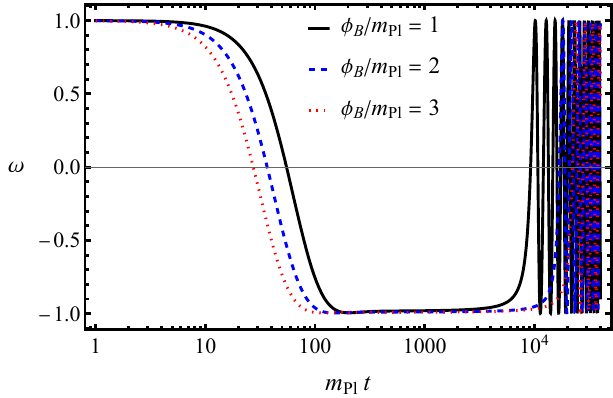}%
}
\caption{\justifying Evolution of the equation-of-state parameter $\omega(t)$
for the string-inspired model in the LQC framework. The slow roll phase concludes as $\omega$ rises above $-\frac{1}{3}$. Each curve corresponds to a distinct initial value of $\phi_{\rm B}$. The initial conditions and model parameters are the same as in {}Fig.~\ref{fig7}.}
\label{fig8}
\end{figure}

\begin{center}
\begin{figure}[!htb]
\subfigure[]{\includegraphics[width=7.5cm]{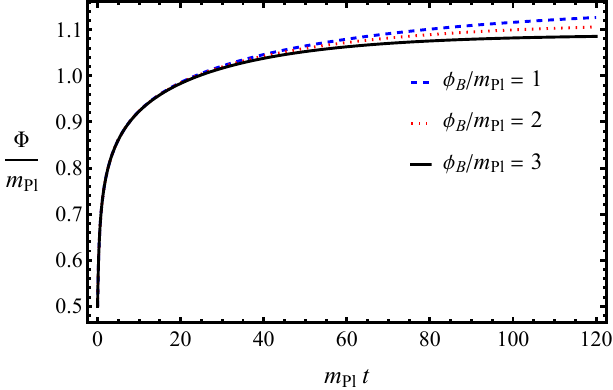}}
\subfigure[]{\includegraphics[width=7.5cm]{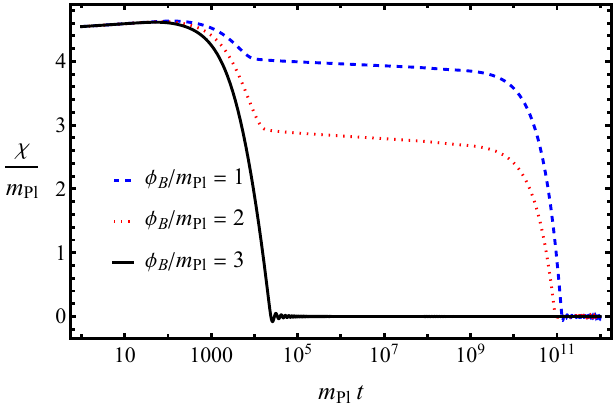}}
\subfigure[]{\includegraphics[width=7.5cm]{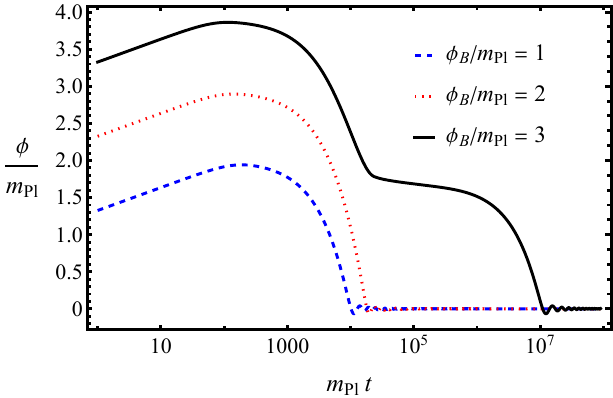}}
\caption{\justifying Evolution for each of the fields, $\Phi(t)$ (panel a), $\chi(t)$ (panel b) and $\phi(t)$ (panel c), for the string-inspired model in the LQC framework. The initial conditions and model parameters are the same as in {}Fig.~\ref{fig7}.}
  \label{fig9}
\end{figure}
\end{center}

\begin{figure*}[!htb] 
\centerline{         \includegraphics[width=1\textwidth]{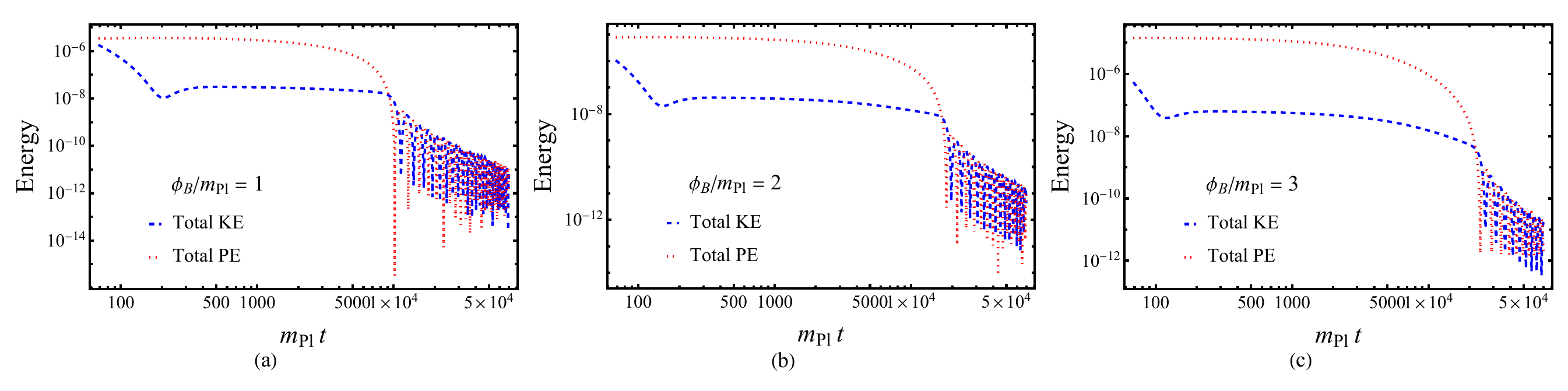}}
    \caption{\justifying Comparison of the kinetic energy and potential energy during the preinflationary and inflationary evolution of the string-inspired model. The initial conditions and model parameters are the same ones considered in {}Fig.~\ref{fig7}, with
    each panel corresponding to a distinct value of $\phi_{\rm B}$. Each panel corresponds to the results for (a) $\phi_{\rm B}/ m_{\rm Pl} = 1$, (b) $\phi_{\rm B}/ m_{\rm Pl}= 2$, and (c) $\phi_{\rm B}/ m_{\rm Pl} = 3$.}
    \label{fig10}  
\end{figure*}

In the string-inspired model Eq.~(\ref{eq:Action_quarticmodel}),
the evolution proceeds through a sequence of dynamical epochs that can be identified consistently across multiple diagnostics. These distinct epochs are summarized in Table~\ref{tab:hm_epochs}.
The resulting background dynamics for the model are presented in {}Figs.~\ref{fig7}--\ref{fig10}.

\begin{table}[htb!]
\caption{\justifying Characteristic epochs (in units of $m_{\rm Pl}^{-1}$) of the quantum-corrected background evolution for the string-inspired model~(\ref{eq:Action_quarticmodel}) with hyperbolic field space geometry, in terms of $\rho$ and $\omega$.  }
    \centering
    \renewcommand{\arraystretch}{1.2}
    \begin{tabular}{lcccc}
    \hline\hline
    &  $\phi_{\rm B} (~m_{\rm Pl}) = $ &$1.0$& $ 2.0$ & $ 3.0$ \\
    \hline
End of superinflation & $\rho = 0.205\,~m_{\rm Pl}^4$ & 0.1798 & 0.1799 & 0.1799 \\
KE$=$PE & $\omega = 0$ & 55.0 & 36.7 & 27.41\\
Onset of slow roll & $\omega \leq -1/3$ & 70.2 & 46.9 & 35.1 \\
End of slow roll & $\omega \geq -1/3$ & 8818 & 16296 & 21400 \\
$\log a_{\text{i}}$ & ... & 2.049 & 1.917 & 1.822 \\
$\log a_{\text{f}}$ & ... & 27.123 & 56.233 & 76.310 \\
\hline
    \textbf{$N_{\rm preinf}=\ln(a_{\rm i}/a_{\rm B})$} & ... & 2.04 & 1.91 & 1.82 \\
\textbf{$N_{\rm inf}=\ln(a_{\rm f}/a_{\rm i})$} & ... & \textbf{25.074} & \textbf{54.315} & \textbf{74.487} \\
\hline\hline
\end{tabular}
\label{tab:hm_epochs}
\end{table}

As shown in {}Fig.~\ref{fig7}, in analogy to the previous studied case for the hybrid model, the Universe evolves from the bounce at $t=t_B$, where $a=1$, $\dot{a}=0$, and enters into exponential inflationary expansion regime. 
The superinflation occurs shortly after the bounce and ends at the epoch listed in Table~\ref{tab:hm_epochs}. The time derivative of the Hubble parameter $\dot{H}$ at this epoch transitions from positive to negative values. When superinflation ends, the energy density $\rho$ reaches half of the critical value, $\rho = 0.205~m_\mathrm{Pl}^4$, at the same epoch in consistence with other indicators.

Upon transition, the Universe enters into the damping phase as evident from the equation-of-state parameter shown in Fig.~\ref{fig8}, where $\omega$ evolves slowly from $1$ towards the inflationary regime with $\omega \sim -1$. The transition epochs for distinct initial values are listed in Table~\ref{tab:hm_epochs}. 
Moreover, these results remain robust under variations of the initial conditions, demonstrating the stability of the background dynamics.

The slow-roll phase is characterized by a sustained dominance of total potential energy~\eqref{stringV} over total kinetic energy, $V(\sigma) \gg \dot{\sigma}^2/2= (\dot{\phi}^2 + e^{2\beta\phi} \dot{\chi}^2 + \dot{\Phi}^2)/2$, nearly constant $H$, and $\omega \approx -1$. Both slow roll parameters $(\epsilon_H, \eta_H) \ll 1$, confirming adherence to slow roll conditions throughout this period. The total number of e-folds for each chosen $\phi_{\rm B}$ value are summarized in Table~\ref{tab:hm_epochs}.

\textcolor{black}{In the equation of motion for $\chi$, Eq.~\eqref{eq:KG_quarticmodel}, the term $2\beta\dot{\phi}\dot{\chi}$ acts as an additional friction or antifriction contribution depending on the sign of $\beta\dot{\phi}$. In this case, for $\beta<0$ and $\dot{\phi}>0$ , e. g., the term effectively behaves as an antifriction term for the $\chi$ field. Consequently, with increase in $\phi_{\rm B}$ more energy is transferred from the $\phi$ sector to the $\chi$ sector, delaying the relaxation of $\chi$ toward its minimum and postponing the onset of oscillations.}

\textcolor{black}{The dynamics for each of the fields are shown in {}Fig.~\ref{fig9} when starting from the bounce. As in the hybrid inflation case, the clock field, with dynamics shown in {}Fig.~\ref{fig9}(a), once again evolves monotonically for this model. Its evolution is sensitive to the chosen initial conditions, which determine the approached constant value asymptotes, similar to the case in the hybrid inflation model discussed in Sec.~\ref{sec:bg_hm}.
The dynamics of the other two fields, $\chi$ and $\phi$, are shown in {}Figs.~\ref{fig9}(b) and ~\ref{fig9}(c), respectively.
Note that although the global inflationary phase terminates precisely when the total Hubble slow roll parameter exceeds unity ($\epsilon_H \equiv -\dot{H}/H^2 > 1$), this condition does not imply that all constituent scalar fields simultaneously exit their respective slow roll regimes. In the context of the present heterotic-inspired model, the global kinetic energy is shared among the fields via the curved target space metric, such that $\epsilon_H = \epsilon_\phi + \epsilon_\chi + \epsilon_\Phi$. If inflation ends due to a rapid condensation or steepening in the $\phi$ (or $\chi$) sector, the individual directional parameter for the geometrically coupled field can remain deeply suppressed (e.g., $\epsilon_\chi \ll 1$, or $\epsilon_\phi \ll 1$). Consequently, if its effective mass satisfies $m_{\text{eff},\chi(\phi)} \ll 3H_{\text{end}}$, the field $\chi\, (\phi)$ can remain heavily overdamped by the residual Hubble friction long after the Universe has transitioned into a radiation-dominated era. It continues to crawl slowly along its local path and only undergoes standard coherent oscillations much later, namely when the Hubble parameter dilutes sufficiently to satisfy the underdamping threshold $3H(t) \lesssim m_{\text{eff},\chi(\phi)}$.
This is the behavior which we see for the dynamics of the $\chi$ field in {}Fig.~\ref{fig9}(b) for the cases of $\phi_B=1 m_{\rm Pl}$ and $2 m_{\rm Pl}$, where inflation ends much earlier, but $\chi$ in these two cases still evolves slowly before finally entering the oscillatory regime much later.
Likewise, we see the opposite we see in the dynamics of $\phi$ shown in  {}Fig.~\ref{fig9}(c), with the cases with $\phi_B=1 m_{\rm Pl}$ and $2 m_{\rm Pl}$ now quickly entering the oscillatory regime right after inflation ends, while the case with $\phi_B = 3 m_{\rm Pl}$ remains for much longer in a slow roll regime.
{}Finally, the plateau change seen in the {}Figs.~\ref{fig9}(b) and ~\ref{fig9}(c) for the evolution of $\chi(\phi)$ at some given time coincides with the time where the other field, $\phi(\chi)$, enters the oscillatory regime and consequent change in the Hubble friction at that instant. }

The evolutions of the energy densities are shown in Fig.~\ref{fig10}.
As expected, the total energy density gradually decreases as the Universe evolves. Initially, the kinetic energy dominates the total energy density, driving the dynamics near the bounce. The large kinetic contributions arise primarily from the initial velocities of the fields, which decrease as the time progresses due to strong Hubble damping. The rapid decay of the kinetic energy is consistent with the observation that the $\chi$ field gradually decays due to its enhanced effective mass. The potential energy is then effectively governed by $\frac{1}{2}m^2 \phi^2$ as the interaction term becomes negligible with $\chi \rightarrow 0$.

{}For larger initial field values, the damping of the potential energy occurs more slowly than the decrease of the kinetic energy, leading to an overall faster approach towards the slow roll regime. Both the kinetic and potential energy contributions show a gradual decline with time before stabilizing at late epochs, as seen in Fig.~\ref{fig10}.

To understand how the preinflationary conditions influence the total expansion, we evaluated $N_{\rm inf}$ as a function of the initial inflaton amplitude $\phi_{\rm B}$ for several representative values of the secondary field $\chi_{\rm B}$. Similarly, we repeated the exercise by adopting the same initial amplitudes for $\chi_{\rm B}$, and evaluated $N_{\rm inf}$ as a function of the initial values of the inflaton $\phi$. 

The $N_{\rm inf}$ corresponding to configurations with positive initial field velocities, i.e., $\dot{\phi}_{\rm B} > 0$, $\dot{\chi}_{\rm B} > 0$  and negative initial field velocities, $\dot{\phi}_{\rm B} < 0$, $\dot{\chi}_{\rm B} < 0$ for which $\sim 60$ e-folds are achieved, are listed in Table~\ref{tab:inf_combined_string_same}.
The number of e-folds first increases, reaches a maximum and then decreases with increasing $\phi_{\rm B}$ for same $\chi_{\rm B}$ value. This behavior is observed consistently for all values of $\chi_{\rm B}$ considered in the initial value scan. For example, in the $(\dot{\phi}_{\rm B},\dot{\chi}_{\rm B})=(+,+)$ configuration with $\chi_B = 7 \, m_{\rm Pl}$, the inflationary duration increases from $N_{\rm inf}\simeq 45$ at $\phi_B = 1.8 \, m_{\rm Pl}$ to $N_{\rm inf}\simeq 93$ at $\phi_B = 3 \,m_{\rm Pl}$, before decreasing again for larger $\phi_B$. A similar pattern is present in the $(\dot{\phi}_{\rm B},\dot{\chi}_{\rm B})=(-,-)$ case.
 {}For extremely small inflaton amplitudes ($\phi_{\rm B} \lesssim 10^{-2}\, m_{\rm Pl}$), the total e-folding saturates at some value that varies with $\chi_{\rm B}$. The same trend is observed when $\chi_{\rm B}$ is varied with same $\phi_{\rm B}$ values.

Additionally, in cases of same sign initial velocity configurations, the values of $\phi_{\rm B}$ for which the the maximum number of e-folds are achieved are higher as $\chi_{\rm B}$ is increased. However, the value of $\phi_{\rm B}$ required to obtain approximately 60 inflationary e-folds decreases as $\chi_{\rm B}$ becomes larger (see Table~\ref{tab:inf_combined_string_same}).
The opposite sign configurations, $(\dot{\phi}_{\rm B},\dot{\chi}_{\rm B})=(+,-)$ and $(\dot{\phi}_{\rm B},\dot{\chi}_{\rm B})=(-,+)$, display the same qualitative behavior of an initially increasing number of e-folds and subsequently decreasing after reaching a maximum value. However, in these cases, the value of $\phi_{\rm B}$ corresponding to $N_{\rm inf}\simeq 60$ shifts towards larger values as $\chi_{\rm B}$ increases (see Table~\ref{tab:inf_combined_string_opp}).

The nonmonotonic dependence of $N_{\rm inf}$ on $\phi_{\rm B}$ arises from a competition between the increasing potential energy available at larger inflaton amplitudes and the corresponding increase in the effective force driving the fields toward the minimum. The hyperbolic field-space geometry further modifies this balance through the factor $e^{2\beta\phi}$, which suppresses the kinetic contribution of the secondary field at larger $\phi$. The maximum inflationary duration is obtained when these competing effects are optimally balanced. Increasing $\chi_{\rm B}$ enhances the interaction contribution $g^2\phi^2\chi^2/2$, thereby increasing the effective restoring force. Consequently, a larger initial inflaton amplitude is required to reach the same balance, causing the location of the maximum $N_{\rm inf}$ to shift systematically toward higher values of $\phi_{\rm B}$.
\begin{table}
\caption{\justifying
Number of preinflationary $N_{\rm preinf}$ and inflationary e-folds $N_{\rm inf}$ for those velocity sign configurations for which enough inflationary e-folds could be obtained for the chosen parameter values in the string-inspired model with a noncanonical hyperbolic kinetic space represented by the metric~(\ref{eq:metric_quarticmodel}), in LQC framework. Highlighted rows correspond to cases with $N_{\rm inf} \sim 60$. The model parameters are $m=10^{-6}\; m_{\rm Pl},\; g= 3\times10^{-4},\; \beta=-0.1 m_{\rm Pl}^{-1}$. The initial values of the clock field and its velocity are fixed at $(\Phi_{\rm B}, \dot{\Phi}_{\rm B})=(0.5~m_{\rm Pl}, 0.5~m_{\rm Pl}^2)$ and $\dot{\chi_{\rm B}}= 0.1~m_{\rm Pl}^2 $.
}
\centering
\resizebox{\columnwidth}{!}{
\begin{tabular}{
l l c c @{\hspace{10pt}}
l c c c
}
\hline\hline
\noalign{\vskip 4pt}

\multicolumn{4}{l}{$(\dot{\phi}_{\rm B},\dot{\chi}_{\rm B})=(+,+)$} &
\multicolumn{4}{c}{$(\dot{\phi}_{\rm B},\dot{\chi}_{\rm B})=(-,-)$} \\

\cmidrule(lr){1-4}\cmidrule(lr){5-8}

$\phi_{\rm B}/ m_{\rm Pl}$ & $\chi_{\rm B}/ m_{\rm Pl}$ & $N_{\rm preinf}$ & $N_{\rm inf}$ &
$\phi_{\rm B}/ m_{\rm Pl}$ & $\chi_{\rm B}/ m_{\rm Pl}$ & $N_{\rm preinf}$ & $N_{\rm inf}$ \\

\hline\\[-4pt]

\textbf{4.7} & \textbf{5} & \textbf{1.675} & \textbf{60.398}  &
3.9 & 4 & 1.974 & 55.837 \\

 4.8& 5 & 1.666 & 59.00  &
4 & 4 & 1.962 & 59.775 \\

 4.9& 5 & 1.660 & 57.65  &
\textbf{4.02} & \textbf{4} & \textbf{1.959} & \textbf{60.538} \\

 5& 5 & 1.655 & 56.34  &
4.1 & 4 & 1.953 & 61.417 \\

\textbf{2.2} & \textbf{6} & \textbf{1.806} & \textbf{60.191}  &
4.2 & 4 & 1.938 & 59.078 \\

 2.2& 7 & 1.759 & 59.441  &
4.3 & 4 & 1.931 & 57.027 \\

2.3 & 7 & 1.750 & 63.18&
3.8 & 6 & 1.840 & 54.192 \\

2.2 & 10 & 1.648 & 57.707 &
3.9 & 6 & 1.827 & 58.111 \\

2.3 & 10 & 1.637 & 61.405 &
\textbf{3.95} & \textbf{6} & \textbf{1.822} & \textbf{60.110} \\

2.3 & 15 & 1.508 & 59.388 &
4 & 6 & 1.817 & 61.607 \\

2.4 & 15 & 1.496 & 63.130 &
4.1 & 6 & 1.807 & 66.316 \\

\hline\hline
\end{tabular}
}
\label{tab:inf_combined_string_same}
\end{table}

\begin{table}
\caption{\justifying
Number of preinflationary $N_{\rm preinf}$ and inflationary e-folds $N_{\rm inf}$ for those velocity sign configurations for which enough inflationary e-folds could be obtained for the chosen parameter values in the string-inspired model with a noncanonical hyperbolic kinetic space represented by the metric~(\ref{eq:metric_quarticmodel}), in LQC framework. Highlighted rows correspond to cases with $N_{\rm inf} \sim 60$. The model parameters are $m=10^{-6}\; m_{\rm Pl},\; g= 3\times10^{-4},\; \beta=-0.1 m_{\rm Pl}^{-1}$. The initial values of the clock field and its velocity are fixed at $(\Phi_{\rm B}, \dot{\Phi}_{\rm B})=(0.5~m_{\rm Pl}, 0.5~m_{\rm Pl}^2)$ and $\dot{\chi}_{\rm B}= 0.1~m_{\rm Pl}^2 $.
}
\centering
\resizebox{\columnwidth}{!}{
\begin{tabular}{
l l c c @{\hspace{10pt}}
l c c c
}
\hline\hline
\noalign{\vskip 4pt}

\multicolumn{4}{c}{$(\dot{\phi}_{\rm B},\dot{\chi}_{\rm B})=(+,-)$} &
\multicolumn{4}{c}{$(\dot{\phi}_{\rm B},\dot{\chi}_{\rm B})=(-,+)$} \\

\cmidrule(lr){1-4}\cmidrule(lr){5-8}

$\phi_{\rm B}/ m_{\rm Pl}$ & $\chi_{\rm B}/ m_{\rm Pl}$ & $N_{\rm preinf}$ & $N_{\rm inf}$ &
$\phi_{\rm B}/ m_{\rm Pl}$ & $\chi_{\rm B}/ m_{\rm Pl}$ & $N_{\rm preinf}$ & $N_{\rm inf}$ \\

\hline\\[-4pt]

2 & 5 & 1.899 & 54.044 &
4.1 & 4 & 1.929 & 64.149 \\

2.1 & 5 & 1.892 & 57.608 &
4.2 & 4 & 1.919 & 67.894 \\

\textbf{2.17} & \textbf{5} & \textbf{1.885} & \textbf{60.177} &
4.3 & 4 & 1.282 & 62.058 \\

2.2 & 5 & 1.880 & 61.293 &
4.4 & 4 & 1.900 & 63.036 \\

2.3 & 5 & 1.870 & 65.100 &
\textbf{4.5} & \textbf{4} & \textbf{1.891} & \textbf{60.924} \\

7.4 & 7 & 1.445 & 63.264 &
4.6 & 4 & 1.880 & 59.006 \\

7.5 & 7 & 1.438 & 61.913 &
4 & 6 & 1.804 & 62.319 \\

\textbf{7.6} & \textbf{7} & \textbf{1.435} & \textbf{60.605} &
7.7 & 6 & 1.539 & 63.450 \\

7.7 & 7 & 1.432 & 59.321 &
7.8 & 6 & 1.534 & 62.002 \\

2.1 & 10 & 1.666 & 54.228 &
\textbf{7.9} & \textbf{6} & \textbf{1.529} & \textbf{60.592} \\

2.2 & 10 & 1.655 & 57.808 &
8 & 6 & 1.524 & 59.221 \\

2.3 & 10 & 1.644 & 61.501 &
8.1 & 6 & 1.520 & 57.893 \\

2.4 & 10 & 1.633 & 65.311 &
8.2 & 6 & 1.515 & 56.603 \\

\hline\hline
\end{tabular}
}
\label{tab:inf_combined_string_opp}
\end{table}

The preinflationary number of e-folds for this model is found to lie in the range $N_{\mathrm{preinf}} \approx  1-2$ across the scanned phase space for all velocity sign configurations. Although this value is larger than that obtained for the first model, it remains relatively small and therefore does not contribute significantly to the overall expansion of the Universe.

The treatment of analytical background solutions for this model is included in the Appendix.

\subsection{Qualitative analysis: Dynamical system approach}\label{subsec:quarticmodel_dsa}

{}Following the methods in Refs.~\cite{Samart:2007xz, Das:2023rat, Ramos:2001zw, Sharma:2018vnv, Shahalam:2017wba}, we define the following dimensionless variables to construct an autonomous dynamical system by using the energy density equation, as follows:
\begin{align}\label{eq:dynvar_quarticmodel}
x_\phi &\equiv \tfrac{\dot{\phi}}{\sqrt{2 \rho_c}}, \quad
x_\chi \equiv \tfrac{\dot{\chi}}{\sqrt{2 \rho_c}}, \quad
x_\Phi \equiv \tfrac{\dot{\Phi}}{\sqrt{2 \rho_c}},\nonumber\\
y_\phi &\equiv \beta \phi , \quad
y_\chi \equiv \tfrac{g \chi}{m}, \quad 
y_\Phi \equiv \tfrac{\Phi}{m}.
\end{align}
Similar to the previous model, we define $z \equiv \tfrac{\rho}{\rho_c}$. Substituting this definition into the modified Friedmann equation~\eqref{eq:FDeff} again produces the constraint Eq.~\eqref{eq:FDdimless}. The constraint obtained by substituting the definitions~\eqref{eq:dynvar_quarticmodel} in the energy density equation, as follows:
\begin{equation}\label{eq:constraintquartic}
z= x_\phi^2 + e^{2 y_\phi}x_\chi^2 + x_\Phi^2 + \frac{m^2}{2 \beta^2 \rho_c} y_\phi^2 (1+ y_\chi^2).
\end{equation}
As discussed in the hybrid model~\eqref{eq:zcompact} analysis, the modified Friedmann equation in LQC restricts the energy density to the bounded interval $0\le z\le1$. 
Consequently, the physical phase space of the dynamical system is not the full unconstrained space of the variables, but a compact submanifold defined by the energy density constraint.

To make this bounded structure explicit, we adopt the same trigonometric compactification~\cite{Boehmer:2014vea} introduced previously in Eq.~\eqref{eq:zcompact}, which automatically enforces the LQC bound on the energy density. 
In terms of this variable, the modified Friedmann equation again splits into two branches corresponding to the expanding ($H>0$) and contracting ($H<0$) phases, with the bounce occurring at $\theta=\tfrac{\pi}{2}$.

Applying this compactification to the string-inspired model, the energy density constraint~\eqref{eq:constraintquartic} becomes
\begin{equation}
\sin^2\theta
=
x_\phi^2
+ e^{2y_\phi}x_\chi^2
+ x_\Phi^2
+ \frac{m^2}{2\beta^2\rho_c}y_\phi^2(1+y_\chi^2).
\label{eq:constraintquartic_theta}
\end{equation}

Thus, the autonomous system evolves on a compact phase-space manifold, and the admissible fixed points must satisfy both the compactified constraints above. 
As in the hybrid model, the trigonometric parametrization provides a global description of the contracting and expanding branches within a single bounded phase space.

We define the dimensionless evolution parameter $\tau \equiv \frac{t}{\beta}$ so that the time derivative transforms as 
$\frac{d}{d\tau}=\beta\,\frac{d}{dt}$.

Using this relation and the definitions in Eq.~\eqref{eq:dynvar_quarticmodel}, the closed autonomous system governing the evolution of the dimensionless variables takes the form
\begin{align}
\frac{dx_\phi}{d\tau}
&=
-3\beta H\,x_\phi
+\sqrt{2 \rho_c}\,\beta^2 e^{2y_\phi}x_\chi^2
-\frac{m^2}{\sqrt{2 \rho_c}} y_\phi
\nonumber\\
&
-\frac{m^2}{\sqrt{2 \rho_c}}y_\phi y_\chi^2,
\end{align}
\begin{align}
\frac{dx_\chi}{d\tau}
&=
-3\beta H\,x_\chi
-2\beta x_\phi x_\chi
-\frac{g m}{\beta \sqrt{2 \rho_c}} e^{-2y_\phi}y_\phi^2 y_\chi,
\end{align}
\begin{align}
\frac{dx_\Phi}{d\tau}
&=
-3\beta H\,x_\Phi,
\end{align}
\begin{align}
\frac{dy_\phi}{d\tau}
&=
\beta^2\sqrt{2\rho_c}\,x_\phi,
\\[6pt]
\frac{dy_\chi}{d\tau}
&=
\frac{\beta g}{m}\sqrt{2\rho_c}\,x_\chi,
\\[6pt]
\frac{dy_\Phi}{d\tau}
&=
\frac{\beta}{m}\sqrt{2\rho_c}\,x_\Phi.
\end{align}

To investigate the asymptotic behavior of the dynamical system, we analyze its fixed points by finding those phase space points where all autonomous variables become stationary.
To investigate the local stability properties of the fixed points, we linearize the autonomous system around each fixed point by computing the Jacobian matrix. The Jacobian encodes how small perturbations evolve in the vicinity of a fixed point, and its eigenvalues determine the nature of stability.

\begin{table*}[t]
\caption{\justifying
Critical points of the string-inspired model in LQC dynamical system. The modified Friedmann equation admits two branches corresponding to $H>0$ (expanding phase) and $H<0$ (contracting phase). An overbar denotes an arbitrary constant value along that direction, indicating a continuous locus of equilibrium points rather than an isolated critical point.
In this model, both branches yield the same fixed-point structure and eigenvalue spectrum.
}
\centering
\label{tab:fixedpoints_quartic}
\renewcommand{\arraystretch}{1.2}
\begin{tabular}{ c c c c}
\hline \hline
Point & Fixed point coordinates & Eigenvalues & Linear stability\\
\hline

\multicolumn{4}{l}{Expanding branch ($H>0$)}\\

$Q_1^{(+)}$ &
$\begin{aligned}
&(x_\phi,x_\chi,x_\Phi,y_\phi,y_\chi,y_\Phi)\\
&=(0,0,0,0,\bar{y}_\chi,\bar{y}_\Phi)
\end{aligned}$
&
$\left\{
0,0,0,0,\;
\pm\,3\times10^{-6} i\sqrt{1+y_\chi^2}
\right\}$
&
Nonhyperbolic center 
\\

\noalign{\vskip 10pt}
\multicolumn{4}{l}{Contracting branch ($H<0$)}\\

$Q_1^{(-)}$ &
$\begin{aligned}
&(x_\phi,x_\chi,x_\Phi,y_\phi,y_\chi,y_\Phi)\\
&=(0,0,0,0,\bar{y}_\chi,\bar{y}_\Phi)
\end{aligned}$
&
$\left\{
0,0,0,0,\;
\pm\,3\times10^{-6} i\sqrt{1+y_\chi^2}
\right\}$
&
Nonhyperbolic center 
\\
\hline \hline

\end{tabular}

\end{table*}

All fixed points obtained in the system correspond to configurations in which the kinetic contributions of all scalar fields vanish, while the potential energy of the $\chi$ sector remains arbitrary but nonzero. Consequently, the total energy density is entirely potential dominated, leading to an effective equation of state $\omega = -1$.

These solutions therefore represent de Sitter--type cosmological regimes. The scale factor evolves exponentially in the expanding branch and contracts exponentially in the contracting branch.

The presence of arbitrary values in $(y_\chi, y_\Phi)$ phase-space coordinates indicates that the equilibrium configurations form a continuous manifold rather than isolated points. This reflects the existence of flat directions in the potential, along which the gradient of the potential vanishes identically. Consequently, the fixed points correspond to degenerate stationary configurations and their stability cannot be determined from linear analysis alone.

The purely imaginary eigenvalues signal oscillatory perturbations around this locus. Thus, the inflationary solution is a center manifold of de Sitter states.
 
\section{Conclusion}\label{sec:conclusion}

We have extended the LQC framework to a generic multifield setting characterized by a nontrivial field-space metric $G_{IJ}$, which allows for kinetic couplings among the component fields. The nontrivial multifield setup  is constructed by incorporating an internal clock field enabling a well-defined relational evolution. The cosmological system thus obtained is canonically formulated in the context of LQG, where GR admits a Yang-Mills–type phase-space structure in terms of the Ashtekar variables. Implementing the holonomy representation enables the quantization of the gravitational sector coupled to the multifield system. The resulting effective description, obtained using the improved $\bar{\mu}$-scheme polymerization yields quantum-corrected equations of motion governing the background dynamics. 

As expected, the robustness of the quantum bounce is confirmed, with the regularization of the classical big bang singularity. Independent of the potential model or initial field amplitudes, the system consistently enters a superinflationary phase driven purely by loop quantum corrections and ends when $\rho=\rho_c/2$, at which point the effect of quantum corrections weakens and the Universe undergoes a transition phase in which Hubble friction rapidly suppresses the kinetic energy of the scalar fields, enabling transition to a potential-dominated slow roll phase.

We then presented an extensive analysis of the background dynamics for the two representative models starting from a KED quantum bounce.
The subsequent dynamics display characteristic postbounce preinflationary and inflationary regimes, which happen for different initial conditions for the fields and their time derivatives.
    
A key outcome of this study is the strong sensitivity of inflationary dynamics to the preinflationary initial conditions, including the relative signs of the field velocities at the bounce for the two example models we have studied. 
These results highlight the intricate interplay between quantum geometric corrections and multifield interactions, and indicate that achieving the observationally required $N_{\rm inf}\sim 50 - 60$ e-folds of inflation in the models considered often necessitates specific parameter choices or finely tuned initial conditions. For both models considered, the contribution of the preinflationary dynamics to the total expansion remains small, typically not exceeding two e-folds.
    
In addition, in our analysis we have found that the choice of different values of the clock field at bounce does not affect the physical results derived for the models. The clock field is simply a spectator field in the present approach that has only  kinetic energy and which quickly dilutes away in the expanding postbounce regime due to its massless nature. 

A comparison between the hybrid and string-inspired models reveals qualitatively different phase-space structures despite their common LQC origin. 
In both systems, the modified Friedmann equation restricts the physical phase space to a compact region, and the presence of the clock field introduces a neutral direction, leading to loci of fixed points rather than isolated equilibria.

In the hybrid model, we found three distinct sets of critical points that arise at $y_\chi=-1,0,+1$. 
The points at $y_\chi=\pm1$ form one-dimensional lines of equilibria parametrized by the clock-field variable $y_\Phi$, and they behave as nonhyperbolic centers with purely imaginary eigenvalues. 
By contrast, the point at $y_\chi=0$ behaves as a saddle, with stability properties that reverse between the expanding and contracting branches. This indicates that the $\chi=0$ configuration acts as a transient or repulsive state, while the symmetry-broken vacua at $y_\chi=\pm1$ correspond to neutral configurations.

{}For the string-inspired model, we have shown that it exhibits a higher degree of degeneracy. 
Instead of isolated lines, it admits a two-dimensional surface of fixed points parametrized by both $y_\chi$ and $y_\Phi$. 
All eigenvalues either vanish or are purely imaginary, implying that the system is entirely nonhyperbolic near these equilibria. Furthermore, the expanding and contracting branches share identical eigenvalue spectra, indicating a symmetric dynamical structure around the bounce.
 
Our results then show that different scalar field interactions and field-space geometry of the two models can lead to qualitatively distinct cosmological phase-space evolutions, even though both models are governed by the same compactified LQC dynamics that ensure bounded trajectories and the avoidance of classical singularities.

{}Future work aims to extend the present analysis by incorporating perturbative dynamics as done for, e.g., the single-field case~\cite{Bojowald:2008gz}.  This would enable the computation of power spectra and spectral indices to confront the model predictions with cosmic microwave background observations. From a theoretical perspective, exploring a broader class of potentials of multifield forms will help assess the universality and robustness of the inflationary features identified here. In parallel, extending the dynamical systems framework to include higher-order stability analysis can reveal nonlinear stability structures. Finally, comparing the evolution across different geometries of the field space, while keeping the rest of the setup identical, can shed light on how field-space curvature and nontrivial metric couplings should influence the quantum-corrected cosmological evolution. Together, these extensions will provide a comprehensive understanding of multifield inflation in LQC that we hope to tackle in the future.

\begin{acknowledgements}

{M.S. acknowledges Mahidol University (Fundamental Fund: fiscal year 2025 by National Science Research and Innovation Fund (NSRF)}; R.O.R. acknowledges financial support through research grants from Conselho
Nacional de Desenvolvimento Cient\'{\i}fico e Tecnol\'ogico (CNPq),
Grant No. 307286/2021-5, and from Funda\c{c}\~ao Carlos Chagas Filho
de Amparo \`a Pesquisa do Estado do Rio de Janeiro (FAPERJ), Grant
No. E-26/200.415/2026. A.W. is partially supported by the U. S. NSF  Grant No. PHY-2308845.
\end{acknowledgements}
\section*{Data Availability}\label{sec:data_availabilty}
There are no publicly available research data
or software supporting this manuscript. Requests for further information or data should be sent to the authors.
\appendix

\section*{Appendix: Analytical solutions to the background dynamics}\label{sec:analytical_solutions}

In this appendix, we present analytical solutions for the background dynamics in LQC following the methods in Ref.~\cite{Zhu:2017jew}. The analysis is performed in successive phases corresponding to dominant physical effects, starting from the bounce and leading up to the slow roll inflationary phase. 

During the bounce and superinflation phase, the dynamics are dominated by the kinetic energy of the scalar field, so that $\rho \simeq \frac{1}{2}\dot{\sigma}^2$ and $V(\phi^K) \ll \frac{1}{2}\dot{\sigma}^2$.  
The Klein-Gordon equation for the average field $\sigma$ reduces to
\begin{equation}
\ddot{\sigma} + 3 H \dot{\sigma} \simeq 0.
\end{equation}

Defining $u(t) \equiv \dot{\sigma}(t)$, we have
\begin{equation}
\dot{u} + 3 H u = 0 \quad \Rightarrow \quad \dot{\sigma}(t) = \dot{\sigma}_{\rm B} \left(\frac{a_{\rm B}}{a(t)}\right)^3,
\end{equation}
where $\dot{\sigma}_{\rm B} = \sqrt{2\rho_c}$ from the bounce condition $\rho_{\rm B} = \rho_c$

Substituting into the effective LQC Friedmann equation
\begin{equation}
H^2 = \frac{\kappa}{3} \rho \left(1 - \frac{\rho}{\rho_c}\right), \quad
\rho(t) = \rho_c \left(\frac{a_{\rm B}}{a(t)}\right)^6,
\end{equation}
and defining the characteristic bounce timescale
\begin{equation}
t_{\rm B}^2 \equiv \frac{1}{3 \kappa \rho_c},
\end{equation}
the analytical solutions for the scale factor and the field are

\begin{align*}\label{eq:bounce_analytical_soln}
a(t) &= a_{\rm B} \left(1 + \frac{t^2}{t_{\rm B}^2}\right)^{1/6}, \quad
\dot{a}(t) = \frac{a_{\rm B} t}{3 t_{\rm B}^2} \left(1 + \frac{t^2}{t_{\rm B}^2}\right)^{-5/6}, \\
H(t) &= \frac{t}{3 (t_{\rm B}^2 + t^2)}, \quad
\dot{H}(t) = \frac{t_{\rm B}^2 - t^2}{3 (t_{\rm B}^2 + t^2)^2}, \\
\rho(t) &= \rho_c \left(1 + \frac{t^2}{t_{\rm B}^2}\right)^{-1}, \quad
\dot{\sigma}(t) = \sqrt{2 \rho_c} \left(1 + \frac{t^2}{t_{\rm B}^2}\right)^{-1/2}, \\
\sigma(t) &= \sigma_{\rm B} + \sqrt{2 \rho_c} \, t_{\rm B} \, \mathrm{arsinh}\left(\frac{t}{t_{\rm B}}\right).
\end{align*}
The above solutions are valid throughout the bounce and superinflation phase, where the kinetic-dominated approximation holds.  

Let $t_c$ denote the transition time where kinetic and potential energies are equal,
\begin{equation}
\frac{1}{2}\dot{\sigma}_c^2 = V(\sigma_c) \quad \Rightarrow \quad \dot{\sigma}_c^2 = 2 V(\sigma_c),
\end{equation}
and let $t_i$ denote the onset of slow roll ($\omega=-1/3$), for which
\begin{equation}
\dot{\sigma}_i^2 = V(\sigma_i).
\end{equation}

Expanding the background in logarithmic time $x \equiv \ln(t/t_c)$ and keeping first-order terms, we write
\begin{align}
\sigma(t) &\simeq \sigma_c + t_c \dot{\sigma}_c \ln\frac{t}{t_c}, \\
\dot{\sigma}(t) &\simeq \dot{\sigma}_c \frac{t_c}{t}, \\
a(t) &\simeq a_c \left(1 + t_c H_c \ln\frac{t}{t_c}\right).
\end{align}

Evaluating at $t = t_i$, we obtain
\begin{align}
\sigma_i &= \sigma_c + t_c \dot{\sigma}_c \ln\frac{t_i}{t_c}, \\
\dot{\sigma}_i &= \dot{\sigma}_c \frac{t_c}{t_i}, \\
a_{\rm i} &= a_c \left(1 + t_c H_c \ln\frac{t_i}{t_c}\right),
\end{align}
and the onset condition gives
\begin{equation}
\frac{t_i}{t_c} = \sqrt{\frac{2 V(\sigma_c)}{V(\sigma_i)}}, \quad
\exp\Big(-\frac{2(\sigma_i-\sigma_c)}{t_c \dot{\sigma}_c}\Big) = \frac{V(\sigma_i)}{2 V(\sigma_c)}.
\end{equation}

{}For a first-order expansion of the potential near $t_c$, we obtain
\begin{eqnarray}
V(\sigma_i) &\simeq& V_c + V'_c t_c \dot{\sigma}_c \ln \frac{t_i}{t_c}, \\
\dot{\sigma}_c &=& \pm \frac{t_i}{t_c} \sqrt{V_c + V'_c t_c \dot{\sigma}_c \ln\left(\frac{t_i}{t_c}\right)}.
\end{eqnarray}
Thus, for the models under consideration, the potential solutions reduce to
\begin{align}
V(\phi_i,\chi_i) &\simeq V(\phi_c,\chi_c) +\nonumber\\ 
&+\partial_\phi V|_c (\phi_i-\phi_c) + \partial_\chi V|_c (\chi_i-\chi_c).
\end{align}

In the previous equations we also have that $\dot\sigma_c$ for each of the models studied here are given by

\paragraph{Hybrid inflation model:}
\begin{align}
\dot{\sigma}_c^2 &= m^2 \phi_c^2 + g^2 \phi_c^2 \chi_c^2 + \frac{\lambda}{2}(\chi_c^2-M^2)^2.
\end{align}

\paragraph{String-inspired model:}
\begin{align}
\dot{\sigma}_c^2 &= m^2 \phi_c^2 + g^2 \phi_c^2 \chi_c^2.
\end{align}

{}For $t > t_i$, the potential dominates and the slow roll conditions
\begin{equation}
\frac{1}{2}\dot{\sigma}^2 \ll V(\phi,\chi), \quad |\ddot{\sigma}| \ll 3 H |\dot{\sigma}|,
\end{equation}
reduce the {}Friedmann and scalar fields equations to
\begin{align}
H^2 &\simeq \frac{8 \pi}{3 m_{\rm Pl}^2} V(\phi,\chi), \\
3 H \dot{\sigma} + \frac{\partial V}{\partial \sigma} &\simeq 0, \quad
\frac{\partial V}{\partial \sigma} = \frac{\dot{\phi}}{\dot{\sigma}} \frac{\partial V}{\partial \phi} + \frac{\dot{\chi}}{\dot{\sigma}} \frac{\partial V}{\partial \chi}.
\end{align}
Integration yields
\begin{eqnarray}
\int_{\sigma_i}^{\sigma(t)} d\sigma \frac{V(\sigma)}{\partial V / \partial \sigma} \simeq - \frac{m_{\rm Pl}^2}{8\pi} (t - t_i),\\
a(t) \simeq a_{\rm i} \, e^{H_{\rm inf} (t-t_i)}, \\
H_{\rm inf}^2 \simeq \frac{8 \pi}{3 m_{\rm Pl}^2} V(\sigma_i).
\end{eqnarray}



\end{document}